\documentclass[apj]{emulateapj}

\lefthead{Ben-Jaffel \& Hosseini}
\righthead{Energetic Atoms in Exoplanet HD209458b}

\def \lya\ {Ly-$\alpha\, $} 
\def \sn\ {{\it S/N}}
\def \bjla  {BJ07}
\def \bjlb {BJ08}
\def \vma {VM03}

\def \vmb {VM04}

\begin{document}

\title{On the existence of energetic atoms in the upper atmosphere of exoplanet HD209458b}

\author{Lotfi Ben-Jaffel \altaffilmark{1}}
\affil{UPMC Univ Paris 06, UMR7095, Institut d'Astrophysique de Paris, 
F-75014, Paris, France; bjaffel@iap.fr}
\altaffiltext{1}{CNRS, UMR7095, Institut d'Astrophysique de Paris, F-75014, Paris, France}
\author{Sona Hosseini} 
\affil{Department of Applied Science, UC Davis, One shields avenue, Davis, CA 95616; sshosseini@ucdavis.edu}

\begin{abstract}
Stellar irradiation and particles forcing strongly affect the immediate environment of extrasolar giant planets orbiting near their parent stars. However, it is not clear how the energy is deposited over the planetary atmosphere, nor how the momentum and energy spaces of the different species that populate the system are modified. Here, we use far-ultraviolet emission spectra from HD209458 in the wavelength range $\left(1180-1710\right)$\AA\ to bring new insight to the composition and energetic processes in play in the gas nebula around the transiting planetary companion. In that frame, we consider up-to-date atmospheric models of the giant exoplanet where we implement non-thermal line broadening to simulate the impact on the transit absorption of superthermal atoms (HI, OI, and CII) populating the upper layers of the nebula. Our sensitivity study shows that for all existing models, a significant line broadening is required for OI and probably for CII lines in order to fit the observed transit absorptions. In that frame, we show that OI and CII are preferentially heated compared to the background gas with effective temperatures as large as $T_{OI}/T_B\sim 10$ for OI and $T_{CII}/T_B\sim 5$ for CII. By contrast, the situation is much less clear for HI because several models could fit the \lya\ observations including either thermal HI in an atmosphere that has a dayside vertical column $[HI]\sim 1.05\times 10^{21}{\rm cm^{-2}}$, or a less extended thermal atmosphere but with hot HI atoms populating the upper layers of the nebula. If the energetic HI atoms are either of stellar origin or populations lost from the planet {and energized in the outer layers of the nebula}, our finding is that most models should converge toward one hot population that has an HI vertical column in the range $[HI]_{hot}\sim \left(2-4\right)\times10^{13}{\rm cm^{-2}}$ and an effective temperature in the range $T_{HI}\sim \left(1-1.3\right)\times10^6$K, but with a bulk velocity that should be rather slow.

\end{abstract}


\section{INTRODUCTION}
The nature of the extended nebula around giant exoplanets orbiting near their parent stars has remained a fundamental unresolved problem since it was raised a few years ago. Of all the exoplanets detected over the last decade, HD 209458b is the first discovered by the transit technique, and the only exoplanet for which rich far-ultraviolet (FUV) space observations were obtained before the Hubble Space Telescope/Space Telescope Imaging Spectrograph (HST/STIS) instrument went down (\cite{vid03, vid04, ben07a, ben08}, hereafter, respectively, \vma, \vmb, \bjla, and \bjlb). The collected data set which covers the wavelength range of 1180-1710 \AA\ is a unique tool not only for understanding the composition and structure of the atmosphere of these giant planets but also for characterizing the processes that heat and excite them. 
Therefore, the interpretation of this data set is of great importance and should ultimately help answer fundamental questions regarding the formation and evolution of extrasolar planets, particularly questions concerning their survival given their close proximity to the harsh environment of their parent stars.

Historically, HD209458b was discovered transiting its parent star and absorbing 1.5\% of its flux (\cite{cha00,hen00}). Extending the technique to space FUV observations at \lya\ (1216 \AA), an enormous hydrogen nebula was discovered using the HST/STIS spectrograph, covering almost $\sim 8.9\%\pm 2.\%$ of the stellar disc (\vma, \bjlb\ ). This corresponds to a hydrogen nebula extending over an effective area that has at least $\sim2.47$ planetary radius across the line of sight, tending to support the theory that the coronal hydrogen gas of HD209458b is inflated. The coincidence between the Roche lobe size across the planet-star line and the occultation effective area suggest that the gas is probably filling the exoplanetary Roche lobe (\vma, \bjlb). Later, based on new HST/STIS FUV observations in the range (1180-1710\AA) of heavy elements like C and O, an occultation, respectively, of $5.\pm2$\% for HI, $13\pm4.5$\% for OI and $7.5\pm3.5$\% for CII were reported (\vmb). There is still a great deal of uncertainty regarding the strong absorptions obtained during transit, respectively, for the two heavy minor constituents, yet the drop-off appears comparable to the transit absorption obtained for HI, which, according to \vmb, supports a blow-off scenario in the upper atmosphere of HD209458b in which an escaping element (HI) drags other heavy constituents outward. The problem is that both OI and CII are several orders of magnitude less abundant than HI, and yet having the two components filling or even overflowing the planet's Roche lobe will not help to attain the reported transit absorptions (\cite{mun07}). In addition, it is not yet clear how {wide} the stellar lines are for the OI and CII multiplets because the HST/STIS observations were obtained in the low-resolution mode that does not help to resolve the emission lines (\vmb, \cite{mun07}). Line broadening processes were thus suggested to explain the OI and CII lines' absorption but they were never implemented in past studies. All these observational, analysis, and modeling uncertainties make the transit absorptions reported for HI, OI, and CII a real challenge for any theoretical model. 

In the following, we revisit available HST archive data in the 1180-1710 \AA\ spectral window to conduct a sensitivity study including HI, OI, and CII abundances together with line broadening processes in play in the upper layers of the atmosphere of HD209458b. Such a global approach is required for any attempt to derive self-consistent observational constraints on the steady-state properties of the nebula around the exoplanet. In Section 2, we recall \lya\ absorption rates and profiles obtained during transit from medium spectral resolutions as reported by {\bjlb\ } and describe in detail the analysis of the low-resolution observations corresponding to HI ($1216$ \AA\ ), OI ($1304$\AA\  triplet), and CII ($1335$\AA\ multiplet). For comparison with our model, a set of four constraints was selected: the absorption line profile at \lya\ as derived from the medium-resolution observations and the line-integrated absorption drop-off during transit corresponding to HI (1216\AA), OI (1300\AA\ triplet), and CII (1335\AA\  multiplet) as derived from the low-resolution observations. Our extended atmospheric model, based on the standard model of \cite{mun07}, is set forth in Section 3 and focuses on the description of the tidal bulge region and stellar emission lines' properties. In Section 4, we describe line-broadening processes that may affect the absorption profiles of the different components. Our results are discussed in Section 5 with the dual aims of deriving a consistent view of the gas distribution and the processes that heat and excite the gas nebula around HD209458b and propounding predictions for future space observations.

\section{Medium and Low Resolution Spectral Observations of HD209458}

Our first data set consists of the HST/STIS archive observation of HD209458 obtained with the G140M medium-resolution grating and the 52"x0".1 slit. The analysis of this data set was controversial, with contradictory conclusions obtained for the \lya\ transit absorption (\vma; \bjla) before the conclusive study in \bjlb\, which we adopt herein. In the global approach that we implement, the transit absorption spectral profile shown in Figure 1 is of particular interest (see also Figure 6 in \bjlb). The symmetric shape of the observed absorption profile, due largely to the coincidence of the transit light curves respectively from the blue and red wavelength regions of the \lya\ line (see Figure 3 in \bjlb), led us to dismiss atmospheric models that produce asymmetric absorption profiles (see \bjlb\ for more details). In addition, considering standard atmospheric models, we previously showed that the hydrogen nebula around HD209458b should be very opaque, behaving much like a damped system than a classical atmosphere. For reference, it is helpful to recall that damped systems are galactic absorbing systems that show extended \lya\ damped wings in the absorption spectra observed toward quasars (\cite{wol86}). For these systems, the [HI] column is usually larger than $2\times 10^{20} {\rm cm^{-2}}$ and the temperature is rather cool (HI thermal velocity $\sim 15 {\rm km s^{-1}}$), similar to the values derived for HD209458b's nebula (\bjlb). However, among all possible optically thick models proposed for HD209458b, no unique solution for the atmospheric hydrogen distribution could be derived as hybrid models with a thin layer of superthermal hydrogen, may also produce satisfactory fits to the observed absorption profile. For those reasons, in the past, we intentionally disfavored any particular model in hopes of obtaining additional observations in the future (\bjlb).
\begin{deluxetable*}{ccccccc}
\tablewidth{7in}
\tablenum{1}
\tablecaption{\small{HST/STIS low resolution data set on HD209458. All observations were obtained with the G140L grating and the 52"x0.2" long slit. Transit central time (TCT) is defined by 2,452,826.628521 HJD (\cite{knu07}). Earth's velocity is projected along line of sight in the heliocentric reference frame.}} 
\tablehead{ \colhead{Dataset name} & \colhead{Program ID} & \colhead{Date Obs.}& \colhead{Time Obs.} & \colhead{Start time - TCT (s)} & \colhead{Duration (s)} & \colhead{Earth's Velocity}}
\startdata
O8U601010 & 10081 & 2003-10-09 & 03:53:41	& -11336.88	& 1780.	& -15.37\\
O8U601020 & 10081 & 2003-10-09 & 05:22:40	& -5997.87	& 2100.	& -15.39\\
O8U601030 & 10081 & 2003-10-09 & 05:22:40	& -237.87   & 2100. & -15.41\\
O8U602010 & 10081 & 2003-10-19 & 19:35:34 & -4436.24	& 1780.	& -19.03\\
O8U602020 & 10081 & 2003-10-19 & 21:11:34 & 1323.76 	& 2100.	& -19.05\\
O8U602030 & 10081 & 2003-10-19 & 22:47:35	& 7084.78	  & 2100. & -19.07\\
O8U603010 & 10081 & 2003-11-06 & 08:44:34	& -10983.48	& 1780.	& -23.66\\
O8U603020 & 10081 & 2003-11-06 & 10:13:18	& -5659.48	& 2100.	& -23.68\\
O8U603030 & 10081 & 2003-11-06 & 11:49:18	& 100.51    & 2100.	& -23.69\\
O8U604010 & 10081 & 2003-11-24 & 00:53:55	& -6709.71	& 1780.	& -26.17\\
O8U604020 & 10081 & 2003-11-24 & 02:20:58	& -1486.73	& 2100.	& -26.17\\
O8U604030 & 10081 & 2003-11-24 & 03:57:00	& 4275.28   & 2100.	& -26.17\\
\enddata
\end{deluxetable*}
The second data set included in this study is the HST/STIS archive observation of HD209458 obtained with the G140L low-resolution grating and the 52"x0".2 slit (\vmb). In this mode, the HST/STIS instrument has a resolution element of $\sim 1.7$ pixels for point sources around \lya\ with an average dispersion of $\sim 0.6$\AA\ per pixel. Generally, 2-3 element resolutions are assumed for the spectral resolution of the instrument. For an extended source, such as the Earth's geocoronal emissions and the sky's background emissions which fill the whole slit area, the spectral resolution is $\sim 8$ pixels with the same dispersion. These details are important because the recorded signal in these data is a combination of a stellar point source emission and the geocoronal and sky background extended sources. The HST/STIS selected mode indeed shows a low-resolution spectrum but covers a larger wavelength band spanning a $\sim 1180-1710$\AA\ window. In the present analysis, we follow exactly the same technique using the time-tag information as described in \bjlb. In total, we have four visits during the planetary transit of HD209458, each composed of three long exposures ($\sim 2000\, {\rm s}$) distributed in time over the transit period (see Table 1). All exposures were obtained with the HST/STIS G140L grating and are sampled here using a time bin of $300\, {\rm s}$ from the available time-tag information. The whole data set resulted in $\sim 67$ bins time series of the HD208458 planet-star system as a function of the orbital phase angle. Coincidentally, the time sampling of the transit period resulted in gaps (lasting, respectively, $\sim 2493\,{\rm s}$, $\sim 1169\,{\rm s}$, $\sim 851\,{\rm s}$, and $\sim 709\,{\rm s}$). The size and number of the gaps, along with the relatively low signal-to-noise ratio (S/N) of the data, make a consistent study of the resulting time series difficult ({\bjla}). Nevertheless, we are able to fit synthetic light curves into each selected wavelength band before we derive the drop-off in the integrated stellar signal during transit, along with the appropriate statistical errors.

\begin{figure}
\epsscale{1.1}
\plotone{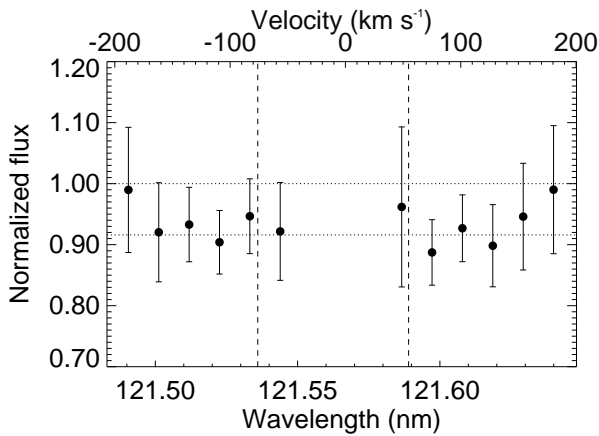}
\caption{Absorption line profile (filled circles) during transit of HD209458b obtained as a ratio between \lya\ line profiles in-transit and unperturbed. The sky background spectral domain as defined here is indicated by two vertical dashed lines. The spectral window was restricted to $\sim \pm 200 {\rm km\, s^{-1}}$ from line center because the signal becomes much too noisy beyond. A flat absorption rate of $\sim 8.4\%$ fits rather well with the obtained absorption profile (dashed line). In \bjlb, we used the DIV1 atmospheric model of \cite{mun07} to propose a Lorentzian-like line profile model with extended wings that also fits quite well with these observations. \label{fig1}}
\end{figure}

For reference, a stellar spectrum is shown in Figure 2. Aside from the emission lines previously described by \vmb\ (their Table 1), a stellar continuum must be subtracted before determining the transit drop-off for individual line or band emissions (e. g., Figure 2A). For that purpose, we use a low-order polynomial model of the logarithm of the stellar continuum spectrum in order to obtain the best fit with the observed spectra in the wavelength window $\sim 1400-1700 $\AA\ (e.g., Figure 2A). After subtraction, we find that no extra correction is needed for any line considered here, which lends support to the continuum model as a good approximation of the stellar continuum (e.g., Figure 2B). The resulting emission lines for the three selected constituents for the in- and out-of-transit periods are shown in Figure 3 with the corresponding statistical errors. Stellar signal drop-off during transit for the emission bands are summarized in Table 2. Our set of results is consistent with \vmb\ within the large statistical uncertainties tied to these observations.

\begin{figure}
\epsscale{1.1}
\plotone{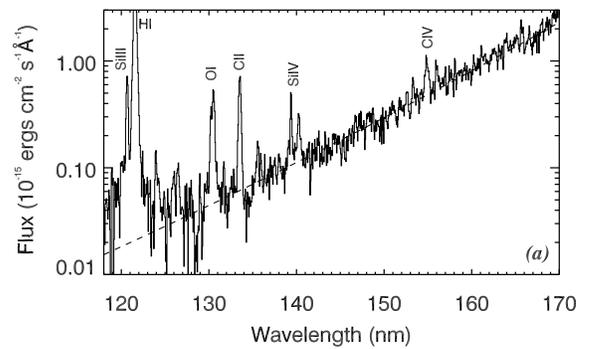}
\plotone{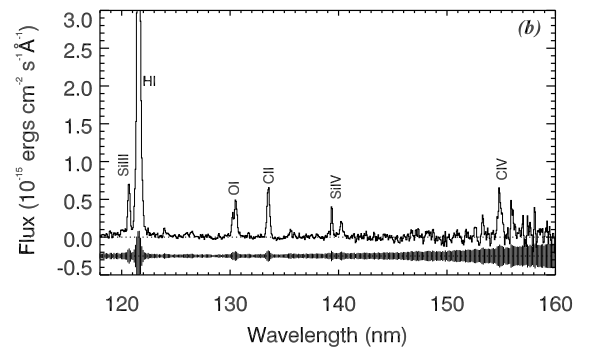}
\caption{(A) Out-of-transit stellar spectrum of HD209458 obtained by HST/STIS G140L low-resolution grating. The time integration corresponds to merging exposures O8U601010 and O8U603010 (e. g., Table 1), similar to \vmb. The stellar continuum shows the classical, solar-like near-UV signal increase trend. The best estimate of the continuum model of HD209458 is over-plotted for reference. (B) Stellar spectrum after subtraction of continuum. Statistical errors are shown for reference.  \label{fig2}}
\end{figure}

In summary, the combination of HST/STIS archive observations results in a set of constraints corresponding to the HI \lya\ medium-resolution absorption profile during transit as shown in Figure ~1, and the transit absorption, respectively, for HI $\sim 6.6\%\pm 2.3\%$ in the wavelength band 1212-1220\AA, OI $\sim 10.5\%\pm 4.4\%$ in the range 1299-1310\AA, and CII $\sim 7.4\%\pm4.7\%$ in the range 1332-1340\AA\ (e. g., Table 2). In the following section, we will define our atmospheric model, the different processes of line broadening at play in the hot atmosphere of HD209458b, and the shape of the stellar emission lines before we conduct a simultaneous comparison to the four constraints assumed here. 


\begin{deluxetable*}{lcc}
\tablewidth{7in}
\tablenum{2}
\tablecaption{Flux drop-off during transit of HD209458 derived from HST/STIS G140L low resolution observations. Typical in \& out transit spectra for the different elements are shown in Fig. 3. (a) The signal drop-off is derived from a best $\chi^2$ fit to the observed time series using synthetic curves for each individual wavelength band. (b) One element is listed but several excited states may also contribute.} 
\tablehead{\colhead{Atoms} & \colhead{Wavelength window (nm)} & \colhead{Flux drop-off $^{a}$ (\%)} }
\startdata
HI & $[121.2,122.0]$          & $ 6.6\%\pm2.3\% $  \\
OI$^{b}$ & $[129.9,131.0]$     &  $ 10.5\%\pm 4.4\% $ \\
CII$^{b}$ & $[133.2, 134.0]$     & $7.4\%\pm4.7\% $  \\
Continuum & $[140.0, 170.0]$     & $1.96\%\pm 0.42\% $  \\
\enddata
\end{deluxetable*}
\begin{figure}
\epsscale{1.1}
\plotone{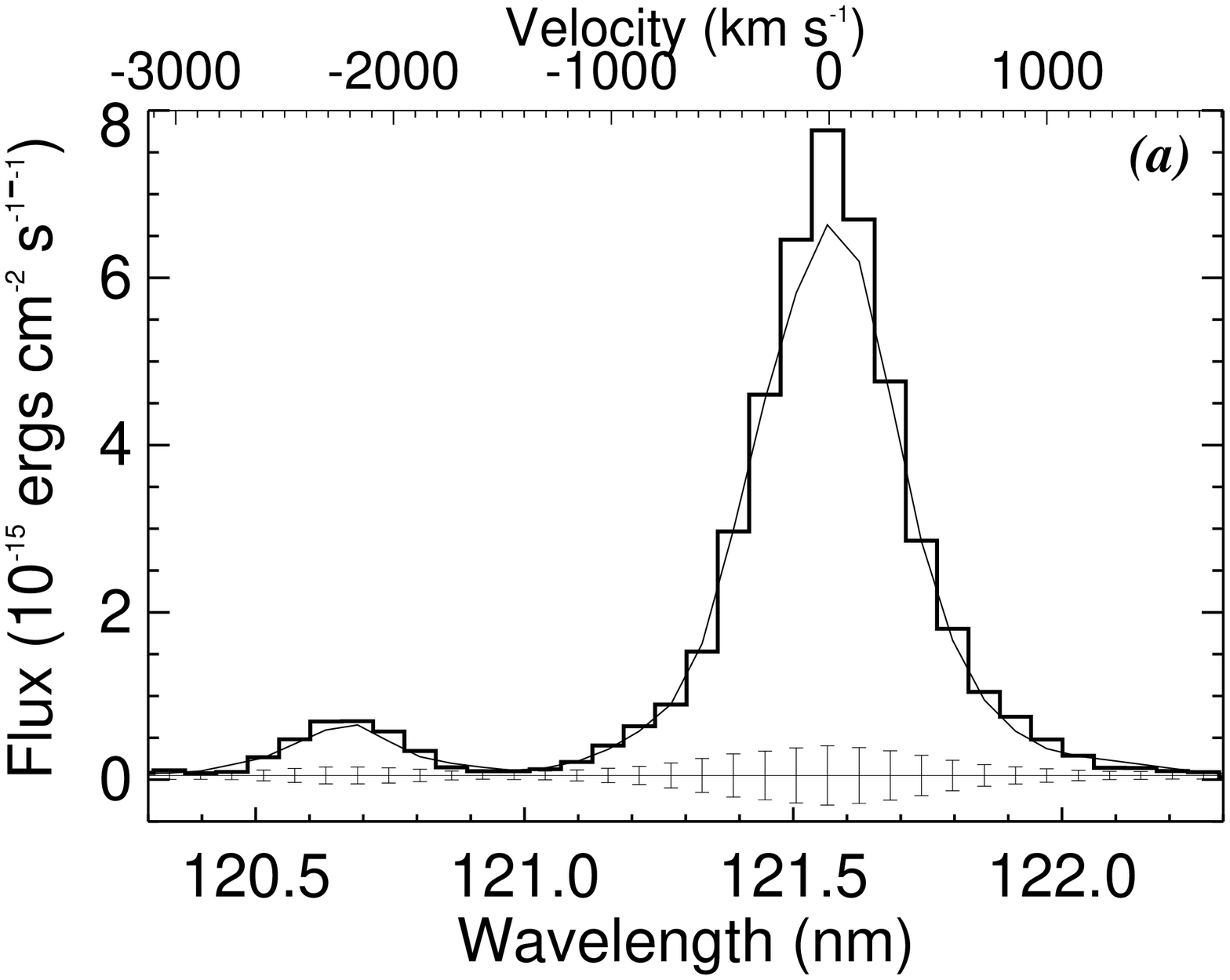}
\plotone{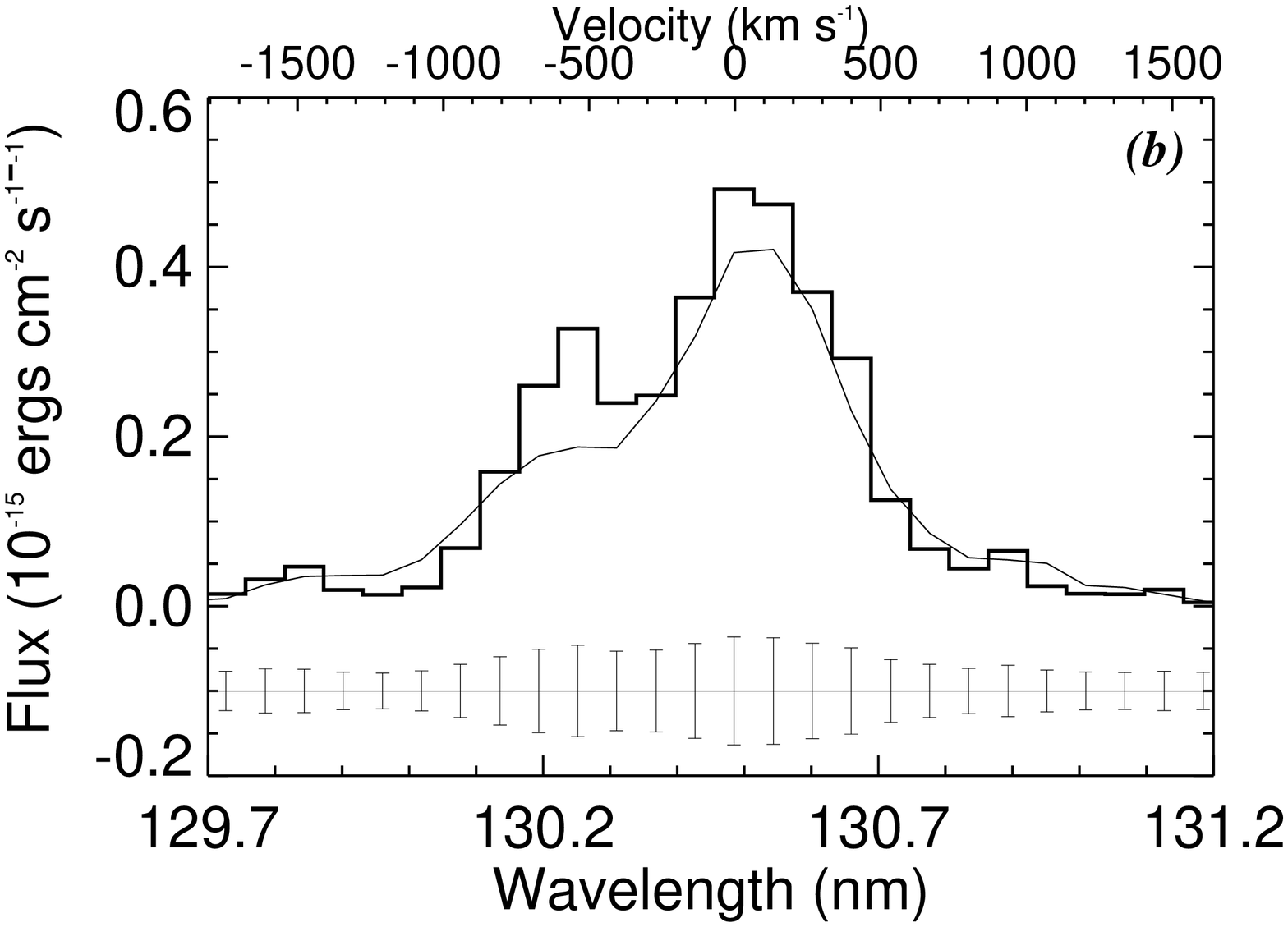}
\plotone{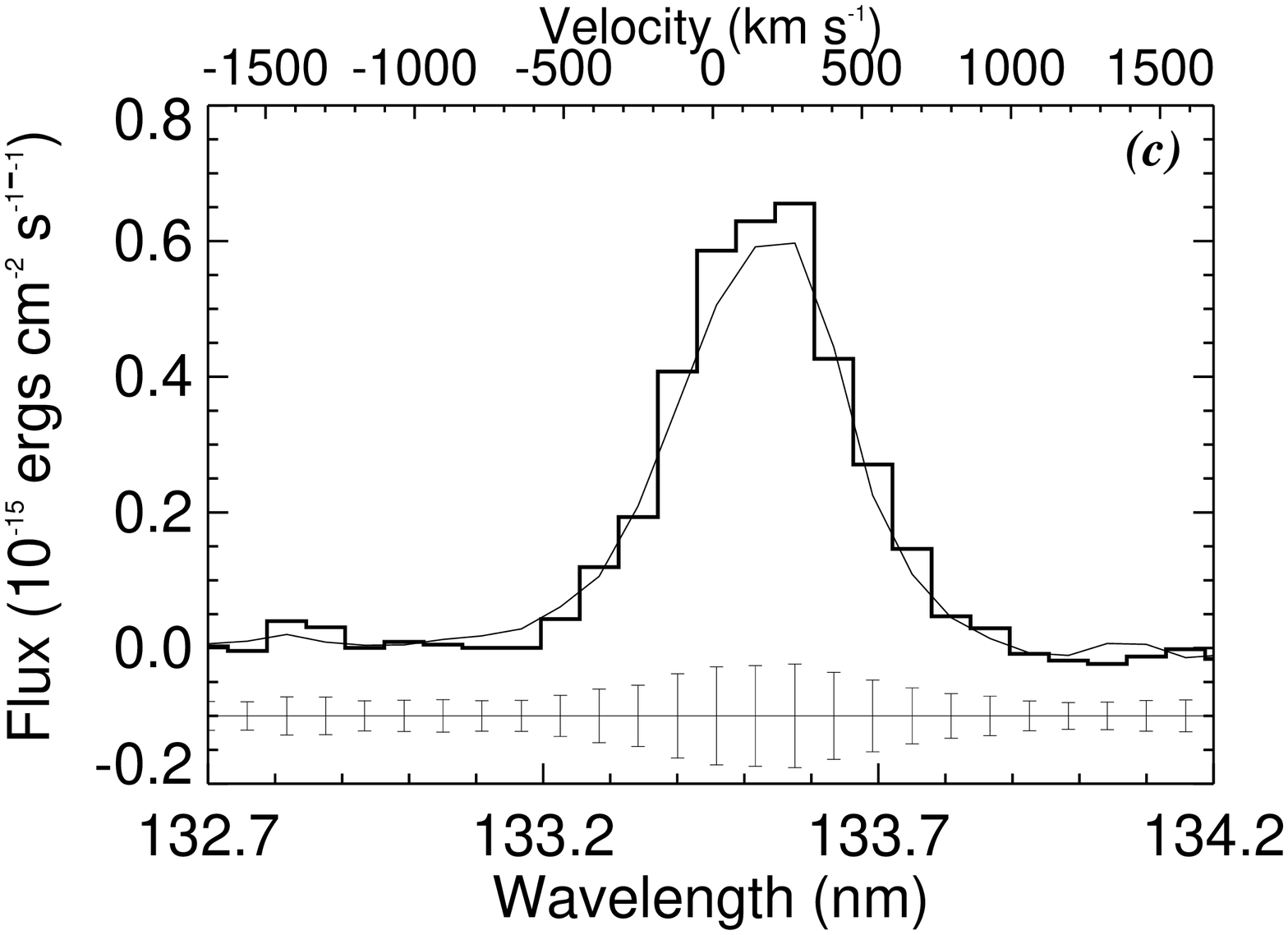}
\caption{Typical in- \& out-of-transit low-resolution line profiles respectively corresponding to HI (a), OI (b), and CII (c). Statistical error bars are shown for the selected in-transit period. \label{fig3}}
\end{figure}

\section{Simulation of Transit Absorption: Method}

In this section, we take a theoretical approach in verifying whether the absorption profile for HI (e.g., Figure 1) and the flux drop-off during transit for the HI, OI and CII lines (e. g., Table 2) are predictable. To evaluate transit absorption spectra, {integration is implemented using a fine grid (see Equation 1 in \bjlb)}, taking into account the varying number density and temperature distributions according to the assumed atmospheric model. We recall that for a target that is not spatially resolved, such as for transiting exoplanets, the observed signal extinction is a weighted average of absorption from different regions of the exoplanetary extended atmosphere, with the outer layers having the maximal weight. In other words, because the planet is not spatially resolved, various sketches of gas distribution around it may lead to the same observed extinction (\bjlb). Ingress and egress observations, which are not yet available with high \sn\ , are required to capture more details of the gas distribution around the exoplanet. In our previous study (\bjlb), we showed that asymmetric solutions in the velocity-space in the example of radiation pressure's induced flow out of the Roche lobe as proposed by \vma\ or energetic neutral atoms (ENAs) produced by the interaction between impinging stellar wind plasma and planetary neutrals (\cite{hol08}) are not compatible with the rather symmetric full-transit absorption as a function of wavelength (\bjlb) if large Doppler-shifted absorptions are assumed. In the example of the ENA model, a slow, external, hot population could be consistent with our data analysis (\bjlb) and will be considered here as a potential absorber of the \lya\ photons during transit. In this study, we thus focus on {cylinder} symmetric atmospheric models with respect to the planet-star line. Any departure from this ideal picture of symmetry that is not detected in our data set could be corrected for afterward starting from the solutions derived in this study. 

\subsection{Atmosperic Model of Exoplanet HD209458b}
In the scenario of atmospheric inflation by stellar radiation and wind particles, exoplanets closely orbiting their parent stars suffer in harsh conditions that may lead to distortion of the outer layers of their atmosphere in unpredictable ways (\cite{yel08,lam09}). For example, the presence of a stellar magnetic field may call for a magnetohydrodynamics (MHD) description of its interaction with a possibly magnetized and out-flowing planet. Depending on the planetary and stellar flow regimes, the strength and relative orientation of the two magnetic fields, and the degree of ionization of the plasma, the final configuration of the interacting system could be very complex, as illustrated by similar situations in our solar system such as the interaction between the solar wind and the local interstellar flow (\cite{rat02}) or the interaction between the Jovian plasma and its satellite Io (\cite{com98, cla02}). The problem is that most models published thus far focus either on a one-dimensional hydrodynamic description that includes chemistry but neglects the interaction with the stellar wind (\cite{yel04, mun07, kos07}), on hydrodynamic simulations that miss most of the underlying micro-physics in the system but include gravity and/or wind interaction with the parent star (\cite{lec04, tia05, sch07, mur09, sto09}), or exclusively on the external wind production of energetic neutrals (\cite{hol08}).

\begin{figure}
\epsscale{1.1}
\plotone{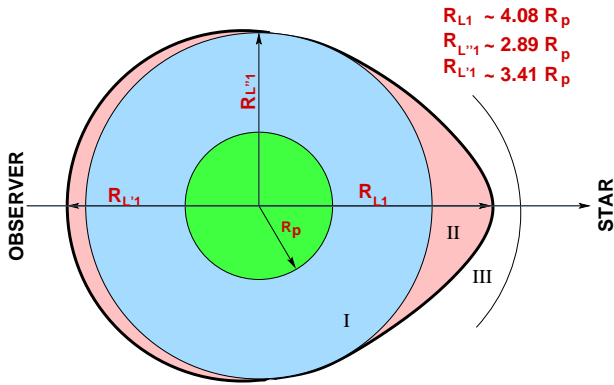}
\caption{Schematic representation of the gas distribution around HD209458b. The planetary gas is assumed to be bound to the planetary Roche lobe (regions I $\&$ II). The corresponding transit absorption is evaluated using the gas opacity inside that boundary. Outside that depicted region, external sources, which may be of stellar origin or populations lost from the planet, form region III, which may exist dayside, nightside, or both. Region III opacity and temperature are free parameters. The spherical region inside $R_{L"1}$ corresponds to the 1D atmospheric model DIV1 of \cite{mun07}. In the bulge region extending from $R_{L"1}$ to the limit of the Roche lobe (the tidal effect), density and temperature are derived from the DIV1 model but may be scaled independently for each specie. This representation sketches most potential gas distributions in the HD209458 planet-star system, including the impact of the stellar wind (see Section 3.1 for more details). \label{fig4}}
\end{figure}

In contrast, this study follows a semi-empirical approach that uses a forward analysis, analyzing simple sketches of the gas distribution in the interaction of the planet-star system and determining how well they fit with the observational constraints. This approach is commonly used and has proven its robustness in several studies of the upper atmosphere of giant planets in the solar system (\cite{gla04, ben07b, sla08}). Our theoretical model is based on sophisticated atmospheric models that include most of the chemistry expected in the HD209458 systems (\cite{mun07}). 
Within that frame, we assume that the exoplanet is filling its Roche lobe with the configuration of gas distribution as shown schematically in Figure 4. The atmosphere is elongated along the planet-star line by the tidal forces (this is the bulge effect). The size of the spherically symmetric atmospheric model (region I) corresponds exactly to the DIV1 model of \cite{mun07} that assumes solar abundances. The main feature of this model is that it includes heavy elements chemistry, which directly enhances the abundance of most species compared to other models, as in \cite{yel04} who only considered hydrogen-helium components. The only change we tolerate for this reference model is scaling the abundance of different constituents over the entire expanse of atmosphere. This scaling should encompass most models thus far published either for the enhancement or depletion of the abundance of the different species. Such abundance variation has its origin in the chemistry thus far assumed, large-scale atmospheric processes as observed in the upper atmosphere of the outer planets of the solar system (\cite{ben93, ben95}, \bjlb), or external sources as suggested by recent two-dimentional hydrodynamic simulation of the planetary wind's interaction with the stellar wind (\cite{sto09}). 

To account for the tidal bulge (region II), we assume the DIV1 model between R$_{L''1}$ (top of region I) and R$_{L1}$ (Roche lobe limit) for the number density and temperature reference distributions. The size of the Roche lobe across the planet-star line should not be much different from the theoretical value of $R_{\rm L''1}\sim 2.89\rm{R_p}$ (\cite{gu03, jar05}). This compares well with the size derived in our previous study of the stellar \lya\ extinction observed during transit (\bjla, \bjlb). The DIV1 atmospheric opacities calculated for the different spectral bands (1216\AA, 1304\AA, and 1335\AA) considered here are shown in Figure 5. For the reference DIV1 model, the dayside vertical column of the three species along the planet-star line, including the bulge region, is $[HI]\sim 1.57\times 10^{21} {\rm cm^{-2}}$, $[OI]\sim 3.94\times 10^{17} {\rm cm^{-2}}$, and $[CII]\sim 2.18\times 10^{17} {\rm cm^{-2}}$.

\begin{figure}
\epsscale{1.1}
\plotone{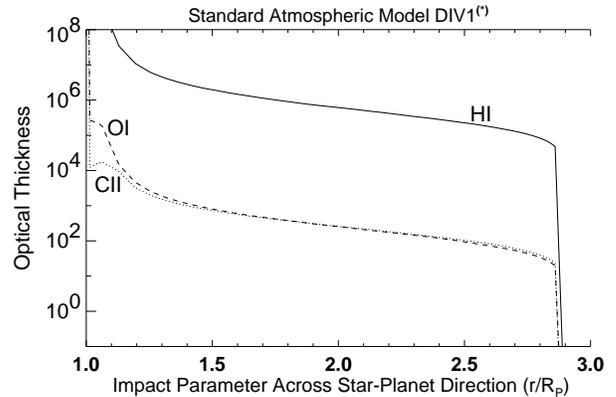}
\caption{Distribution of the planetary atmosphere opacity as a function of the impact parameter measured across the planet-star line, respectively, for HI, OI, and CII. The distribution corresponds to the DIV1 spherical model of \cite{mun07} before any atmospheric scaling is applied. The opacity from the bulge region and/or from any external source should be added for the final estimation of the transit absorption.    \label{fig5}}
\end{figure}

In addition to the thermal population described above, we consider two sources of hot atoms that are included independently in our transit absorption calculations. First, to account for superthermal populations within the upper layers of the nebula (the internal source of hot atoms), two free parameters are introduced, corresponding to the position of the bottom of that active region in the atmosphere measured from the top of region II and the temperature of the hot atoms of the considered specie. Outside of the Roche lobe configuration sketched in Figure 4, we also include the possibility of an external source (region III) of absorbing atoms, described by its own atomic column and temperature. This source could be of stellar origin or populations lost from the planet  dayside, nightside, or both. If the external source is of stellar origin, then it coincides with the ENA model proposed by (\cite{hol08}). Along with the scaling factor for the whole atmosphere, this description should cover most scenarios of gas distribution, including thermal and non-thermal populations as well as those with or without the presence of a stellar wind.
 
In the following section, in order to generate the transit flux drop-off, we evaluate the line profile of the different stellar emissions either directly from observed, unperturbed stellar lines of HD209458 or from available solar emission lines.

\subsection{Stellar Unperturbed Emission Line Profiles}

HD209458 is G0 V solar-like star (\cite{woo05}) in which a large number of emission lines originate from the chromosphere/corona. For \lya\ , this is confirmed by several STIS high/medium resolution observations that show a self-reversal (double horn) shape that has $\sim 1$\AA\  line's width, almost the same properties known for the solar \lya\ line (\vma, \bjla, \cite{lem05}). In addition, we recall that chromospheric CaII H and K lines were observed for HD209458, and these are comparable to solar lines (\cite{shk05}). If we also consider the comparable size, mass, effective temperature, and age of the two stars (\cite{cha00,hen00,mel06,nor04}), all of these factors tend to support the conclusion that the OI and CII solar line profiles are good estimates for HD209458 lines.

In the specific case of the HI \lya\ emission line, the HST/STIS medium-resolution observations of HD209458 provide a good reference for the stellar line profile (see Fig. 3 in \bjla). The so-called unperturbed line profile can be used not only for the analysis of the medium-resolution observations, as done in \bjla, but also can be convolved with the instrument line spread function of the HST/STIS G140L grating in the 52"x0".2 slit mode in order to generate a reference \lya\ line profile for the low-resolution data analysis. In order to verify the accuracy of this method, one may start from high-resolution observation of HD209458 obtained with the HST/STIS echelle mode but at low \sn\ $\,$ (\vma). After convolution with the appropriate line-spread function corresponding to the G140M grating, we confirm that line profiles obtained either directly or after convolution are in agreement. 

In the case of the OI and CII emission lines, the situation is more complicated because the lines' profiles are not fully available. With the G140L spectra, obviously we have no way to access the genuine stellar lines because of the poor spectral resolution. Until recently, it was impossible to state the origin of the OI and CII absorption during transit with any certainty because so much about the stellar line profiles that compose the OI and CII multiplets was unknown. However, we believe the problem can be solved in two steps: first, we derive the weights of the different lines that compose each of the multiplets using available HST/STIS high-resolution spectra of HD209458 (\vma). The data are noisy but the collected spectra allow one to make a good estimate between the lines' peaks in the ratio 1:1.5:1.17, respectively, for the OI ($1302.17$\AA, $1304.86$\AA\ and $1306.03$\AA) lines, and in the ratio 0.5:1.0 for the CII ($1334.53$\AA, $1335.71$\AA) lines. Note that the CII $1335.66$\AA\ line is dismissed because it almost merges with the $1335.71$\AA\ line while having a very weak oscillator strength (e. g., Table 3). 
Because HD209458 is a solar-like star, the second step in modeling the OI and CII multiplets is to determine whether solar spectra of the different OI and CII lines are available. Fortunately, OI $1302.17$\AA, OI $1304.53$\AA, and OI $1306.71$\AA\ line profiles have been measured by the OSO-8 satellite with good accuracy (\cite{lin88}). In addition, line profiles of, respectively, CII $1334.53$\AA\ and CII $1335.71$\AA\ were recorded by a NASA Aerobee sounding rocket (\cite{ber70}) and more recently by the Solar and Heliospheric Observatory/Solar Ultraviolet Measurement of
Emitted Radiation (SoHO/Sumer) spectrometer (\cite{cur01}). The average FWHM of the OI and CII lines are $\sim0.2$\AA\ (or $\sim 45 {\rm km\ s^{-1}}$) and $\sim0.22$\AA\ (or $\sim 50 {\rm km\ s^{-1}}$), respectively. It is important to note that all line profiles have FWHMs far larger than the thermal Doppler width in the HD209458b atmosphere and much larger than previously estimated (\vmb; \cite{mun07}).  In the following section, we show that our finding unquestionably confirms the need for an efficient line broadening in order to explain the strong absorption observed during transit of HD209458. 

\begin{figure}
\epsscale{1.1}
\plotone{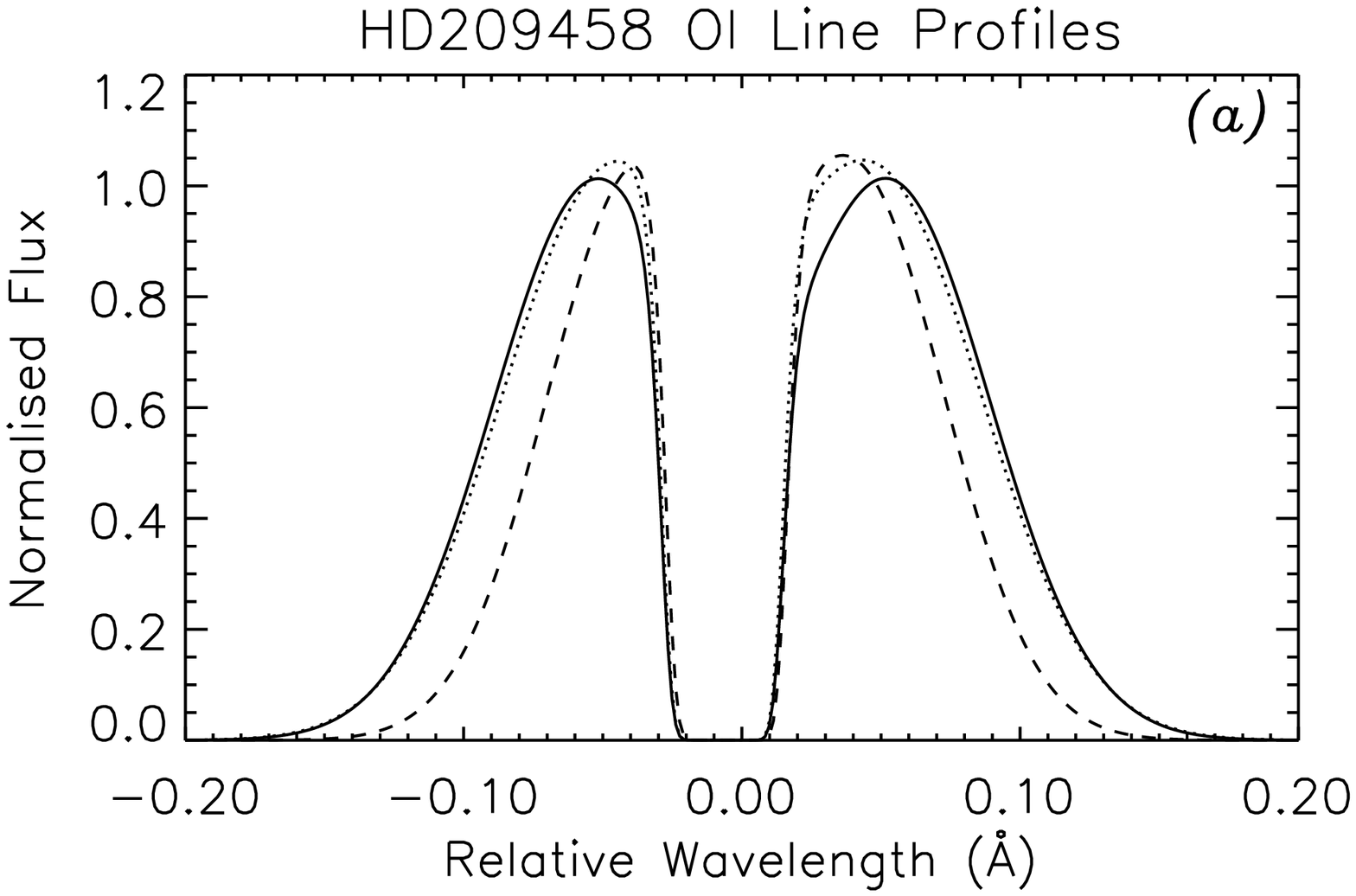}
\plotone{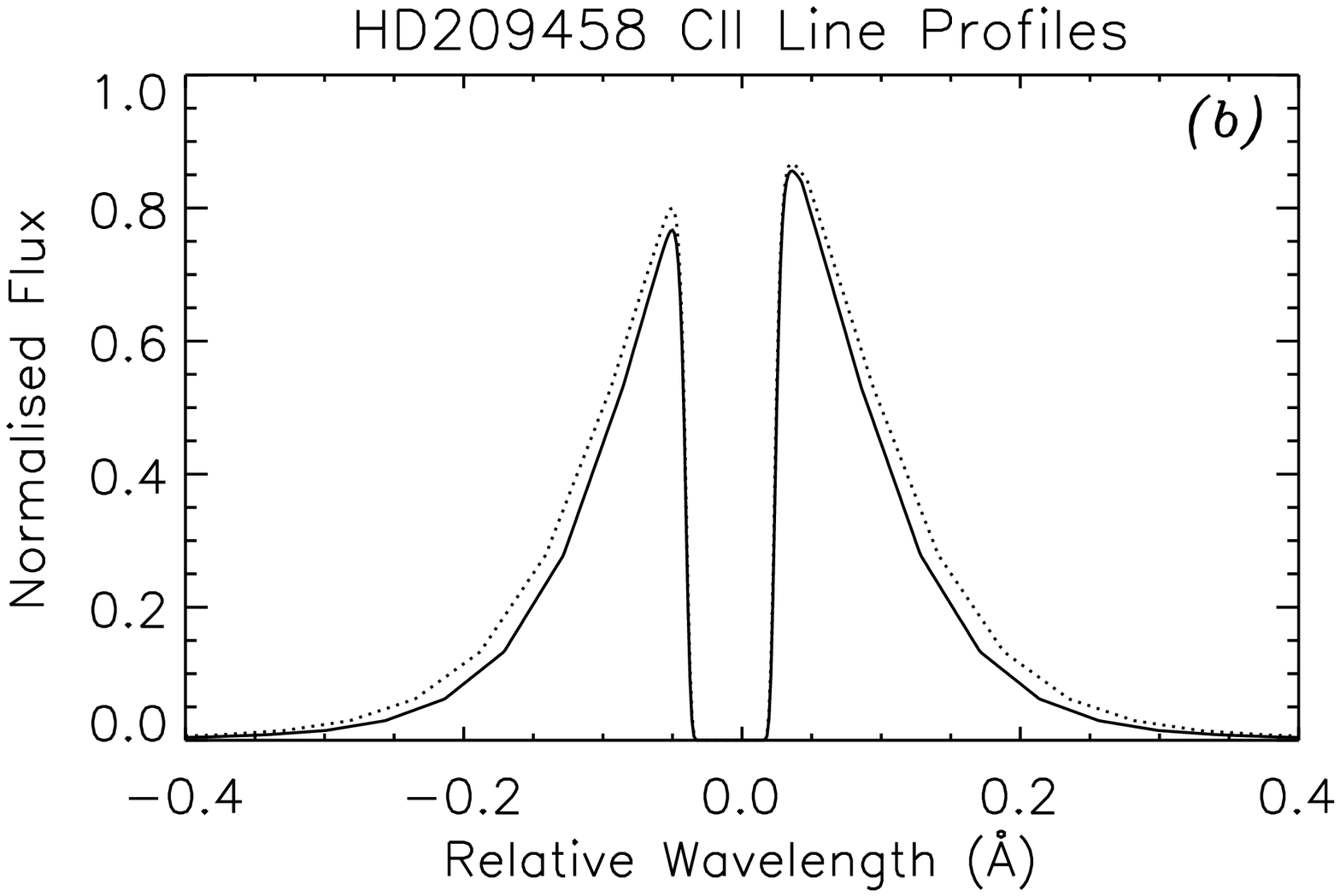}
\caption{Observed solar OI (a) and CII (b) lines assumed here for HD209458 after absorption by the interstellar medium gas between HD209458 and Earth. Most lines have an average width of $\sim 45 {\rm km s^{-1}}$ for the OI triplet and $\sim 50 {\rm km s^{-1}}$ for the CII multiplet, much larger than the Doppler width at the atmospheric temperature of $1.2\times 10^4 {\rm K}$.   \label{fig6}}
\end{figure}

To obtain the stellar OI and CII line profiles at Earth's position, we start with the solar line profiles to which we apply the interstellar medium (ISM) absorption between HD209458 and the observer. As no observation exists for the local interstellar medium (LISM) OI absorption in that direction, we scale from the column obtained for HI (\cite{woo05}) using the $[OI/HI]_{ISM}\sim 10^{-4}$ ratio in the ISM  (\cite{moo02}). For the CII ion column, we use the CII ion column measured toward Capela that we scale for the HD209458 distance (\cite{woo97}). The corresponding profile of the unperturbed stellar OI and CII lines at Earth's orbit are shown in Figure 6. The OI and CII multiplets are built using the peak weights (see above) between the different components as derived from the HD209458 spectrum obtained by HST/STIS in the echelle mode. After the lines are convolved with the appropriate line spread function, a comparison to the low-resolution, unperturbed line profile helps, if needed, to adjust the scaling factor between the different components that make the multiplets. For each individual transition, a Voigt atomic absorption profile is assumed. Corresponding key parameters required to estimate atmospheric opacities and corresponding absorption lines are provided in Table 3.

\begin{deluxetable*}{lccccccc}
\tablewidth{7in}
\tablenum{3}
\tablecaption{\small{Properties of HD209458 stellar FUV Emission Lines. These are the most important lines and continuum considered in the present study. Atomic parameters are provided at a reference temperature of $10^4$K. $ \sigma_0$ is the resonance cross section at line center, ${\rm a_v}$ is the Voigt parameter required to evaluate the atomic absorption profile, and the pair ($x_0$, $x_1$) are the parameters that better represent the stellar line as a double Gaussian profile $\sim \left( \exp{-\left({{x-x_1}\over x_0}\right)^2}+ \exp{-\left( {{x+x_1}\over x_0}\right)^2} \right)$  (\cite{lin88}). The parameter $x$ is the photon's frequency in Doppler units, and ($x_0$, $x_1$) are provided in Doppler units at a reference temperature $10^4$K. The double Gaussian profile only applies for those emission lines that exhibit a double horn shape before absorption by the ISM gas. The OI triplet is due to the transition $3s$ $^3S \rightarrow 2p$ $^3P$. The CII multiplet  results from the transition 2s2p$^2$ $^2D\rightarrow$ 2s$^2$ 2p $^2P^0$.}}
\tablehead{\colhead{Atoms} & \colhead{Wavelength (nm)} & \colhead{A$_{\rm ul} ({\rm s^{-1}})$} & \colhead{$f_{ul} $} & \colhead{$ \sigma_0\,{\rm cm^{2}}$} & \colhead{${\rm a_v}$} & \colhead{$x^{§}_0$} & \colhead{$x^{§}_1$}} 
\startdata
HI    &  121.567        & $6.265 10^8$  & 0.4162 &  $5.88 10^{-14}$ & $4.699 10^{-4}$ & 4.02 & 4.22 \\
OI    &  130.217         & $3.15 10^8$   & 0.048  &  $2.898 10^{-14}$ & $1.012 10^{-3}$ & 3.67 & 3.80  \\
OI    &  130.486         & $1.87 10^8$   & 0.0478  &  $2.892 10^{-14}$ & $6.02 10^{-4}$ & 3.87 & 3.46  \\
OI    &  130.602         & $6.23 10^7$   & 0.0478  &  $2.894 10^{-14}$ & $2.01 10^{-4}$ & 3.24 & 2.84  \\
CII   &  133.453         & $2.4 10^8$   & 0.128  &  $6.859 10^{-14}$ & $6.845 10^{-4}$ & --- & ---  \\
CII   &  133.566        & $4.8 10^7$   & 0.0128  &  $6.859 10^{-14}$ & $1.37 10^{-4}$ & --- & ---  \\
CII   &  133.571         & $2.88 10^8$   & 0.115  &  $6.168 10^{-14}$ & $8.222 10^{-4}$ & --- & ---  \\
Continuum & 140.0-170.0  &     -          &   -      &         -          &       -          &  -   & - \\
\enddata
\end{deluxetable*}

\section{Sensitivity of Transit Absorption to Atmospheric Parameters and Processes}

     \subsection{Thermal Processes: Classical Approach}
     
In the first attempt to compare observations to the transit extinction predicted by our basic model, we consider thermal broadening as the only process responsible for the formation of the atomic absorption profiles. Using the DIV1 atmospheric model of \cite{mun07}, the calculated transit \lya\ absorption profile is slightly wider than the observed profile with a $\chi^2\sim1.98$ (Figure 7A), while the line-integrated transit drop-off at \lya\ attained by the model $\sim 7.4\%$ is also slightly larger than the observed drop-off ($6.6\%\pm2.3\%$), yet still within the measured error bars (see Table 2). The $\chi^2$ indication is based on the medium- and low-resolution \lya\ observations and is provided merely to simplify the comparison between the multiple solutions. One way to decrease the atmospheric extinction is to deplete the gas abundance in the DIV1 model by a scaling factor over all of the atmosphere. For the sake of clarity, we may consider the model proposed by \cite{yel04} and obtain its enhancement in order to fit the observed transit absorption. As described in \bjlb, the origin of such enhancement in the atomic hydrogen abundance could be a different chemistry that includes heavy elements (\cite{mun07}), HI production by collisional processes between magnetospheric particles and the background atmosphere similar to conditions in the auroral regions of Jupiter (\cite{ben93}), or gas accumulation from the stellar wind that may confine the atmosphere (\cite{sto09}). In the example of Saturn, the origin of the HI excess is attributed to HI chemistry enhancement caused by precipitating water by erosion from the extended rings of the planet (\cite{ben95}). The atmospheric enhancement approach was considered in \bjlb, yet it is only relative as one may start from an HI-rich atmospheric model (DIV1) and consider cases of abundance depletion as we do here.

Starting from the DIV1 model, the amount of depletion of the HI abundance is obtained by the application of scaling factor $f_{sca}$ that modifies the final transit absorption profile at \lya\ and the corresponding signal drop-off. As shown in Figure 7A, the thermal broadening of the \lya\ absorption line follows a classical curve of growth compared to the HI column. For reference, for $f_{sca}=1/10$, $1/3$, $2/3$, and $1$ cases shown in Figure 7A, the corresponding $\chi^2$ is, respectively, $\sim 5.68$, $\sim 2.42$, $\sim 1.73$, and $\sim 1.98$. The best fit for the transit absorption profile is obtained for a scaling factor $f_{sca}\sim 2/3$ with a $\chi^2=1.7289$. A transit drop-off of $\sim 6.74\%$ of the stellar flux integrated in the wavelength range $\sim 1212-1220 $\AA\ consistently fits quite well with low-resolution observations $\sim6.6\%\pm2.3\%$ (e. g., Table 2). This family of solutions based on scaling the HI abundance was initially proposed in our previous study (\bjlb) comparing the gas nebula around HD209458b to damped systems. However, one of our conclusions in \bjlb\ emphasized that no unique solution could be favored to explain the extinction level observed in the HI \lya\ line profile and integrated flux. This multiplicity of solutions is the reason why \bjlb\ chose not to disseminate quantitative details concerning any particular solution until the present in-depth study.

\begin{figure}
\epsscale{1.1}
\plotone{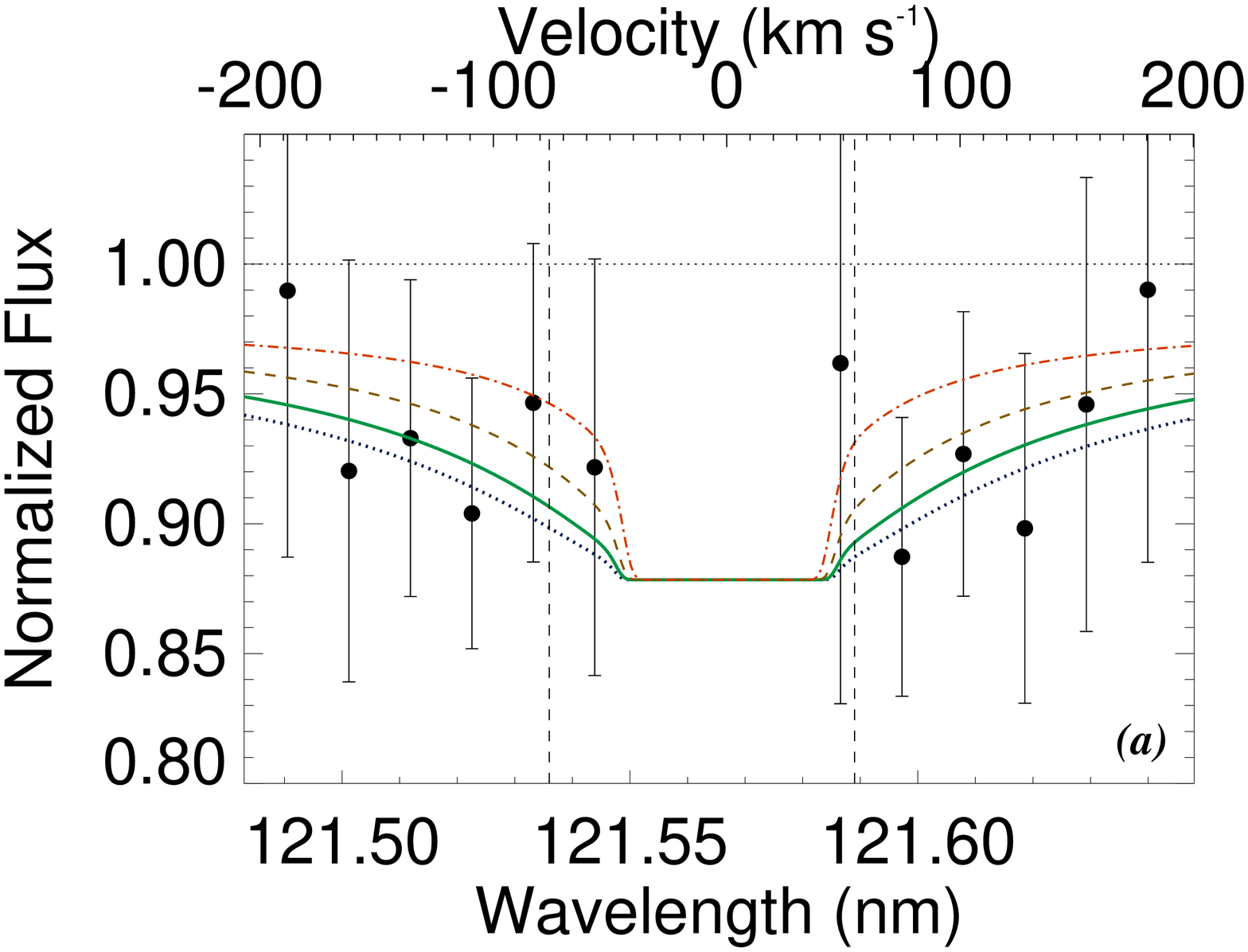}
\plotone{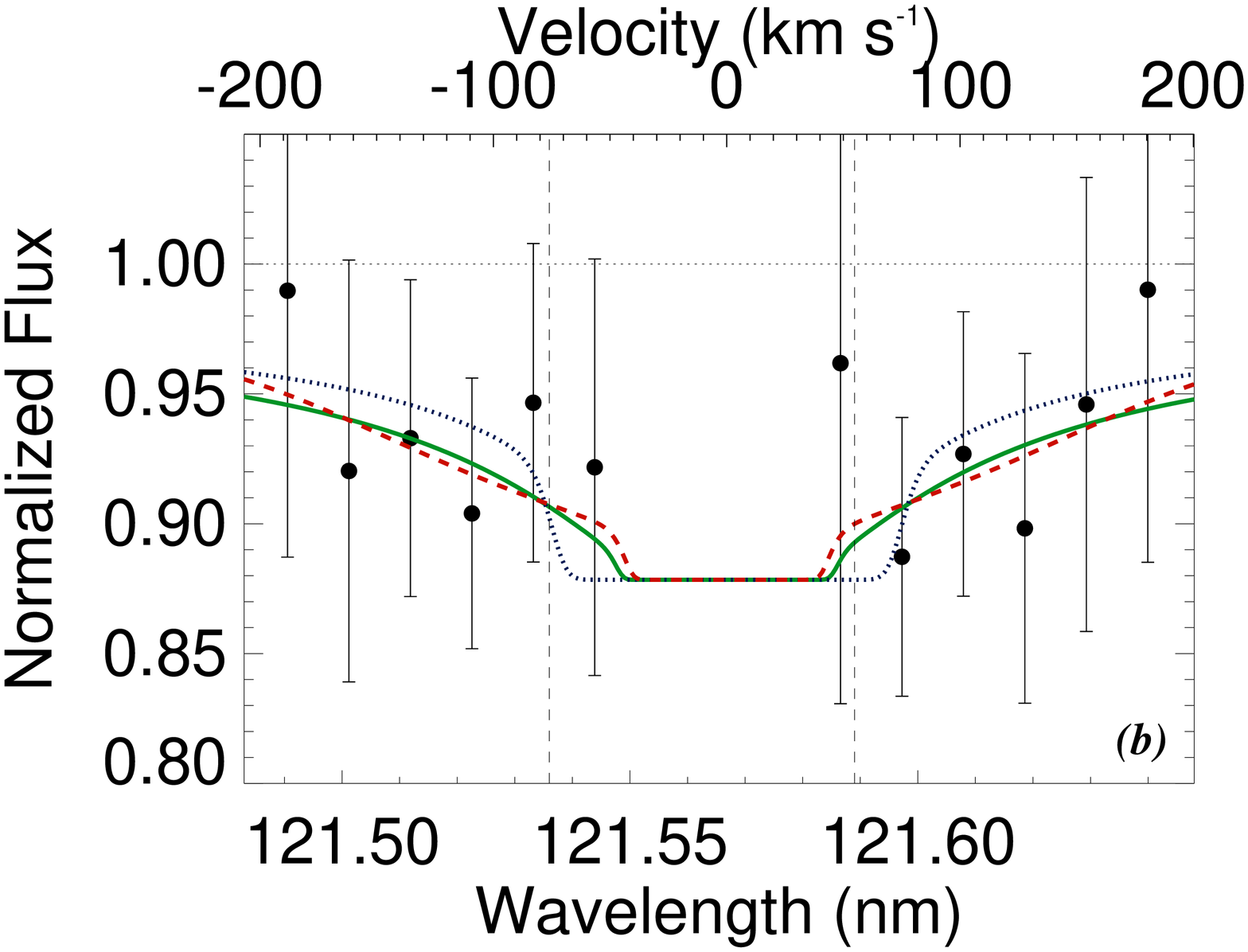}
\caption{(a) Comparison of observed \lya\ absorption line profile (filled circles) during transit of HD209458b to theoretical models assuming only thermal broadening as a function of the atmospheric scaling factor $f_{sca}$. (Dotted line) $f_{sca}=1$, (solid line) $f_{sca}=2/3$ represents the best fit in terms of $\chi^2$ for these kinds of models, (dashed line) $f_{sca}=1/3$, and (dotted-dashed line) $f_{sca}=1/10$. (b) Comparison between observed (filled circles) and theoretical best fit \lya\ absorption line profile during transit of HD209458b obtained for: (solid line) atmosphere scaled by $f_{sca}=2/3$ with thermal broadening alone, (dotted line) atmosphere scaled by $f_{sca}=1/3$ with energetic HI ($T_{HI}/T_B=3.5$) in region II, and (dashed line) atmosphere scaled by $f_{sca}=1/10$ with energetic HI confined to region III ($[HI]_{hot}=4.\times10^{13} {\rm cm^{-2}}$ and $T_{HI}/T_B=102$). Note the spectral inflection that appears at $\sim \pm 100$km/s in the wings of the \lya\ absorption profile when superthermal HI atoms are included.  \label{fig7}}
\end{figure}

The next logical step is to determine how the atmospheric model tested in the HI \lya\ analysis would work for the OI triplet and the CII doublet. For that reason, we start from the DIV1 model using the gas configuration sketched in Figure 4 and calculate the total opacity as a function of distance from the planet's center, across the planet-star line. As for HI, the OI and CII abundances can be independently scaled by a factor over the whole atmosphere. We assume that all lines that compose the multiplets also contribute to the absorption of the stellar flux. Indeed, for OI, the $1302.17$\AA\ line is strongly absorbed by the ISM, making it the largest contributor to the unresolved multiplet signal from the two remaining lines ($1304.86$\AA\ and $1306.03$\AA\ ) that should be collisionally excited (\vmb). For CII, the observed multiplet signal should originate respectively from CII $1334.53$\AA\ and CII$^*$  $1335.71$\AA\, the latter being collisionally excited. With the opacity of OI and CII shown in Figure 5, it seems quite difficult for the DIV1 model to fit the extinction observed for OI and CII during transit even after scaling by any factor considered for HI. For reference, for $f_{sca}=2/3$, the line-averaged flux drop-off during transit for the OI triplet is $\sim 3.9\%$, and $\sim3.3\% $ for CII. As shown in Table 2, these values fall in the bottom range observed for the transit absorption, making thermal broadening in scaled atmospheric models a marginal solution for the OI and probably CII lines. 

The challenge we face with OI and CII transit absorptions is twofold. Indeed, assuming the transit absorption shown in Table 2, and despite the large statistical uncertainties, how is it that minor constituents that are 4 orders of magnitude less abundant than HI show similar, if not stronger, absorption during transit? In addition, the transit absorption seems even stronger for neutral OI than for ion CII. It follows that for the OI and CII absorption lines in the HD209458b extended atmosphere, another explanation is required that may compensate for their weaker abundances compared to HI and that might then provide comparable absorption drop-off during transit for the three constituents. The missing mechanism that may explain the high absorption observed for both OI and CII lines may be related to broadening processes that should be in play in the upper atmosphere of HD209458b. This was discussed qualitatively in the past (\cite{mun07}) but no attempt was made to evaluate the spectral broadening that may explain the large absorption observed during transit. In addition, the unknown spectral shape of the stellar lines at selected wavelengths was a source of great uncertainties on modeling the transit absorption of OI and CII. In the following, we elicit key information on the shape of the stellar lines and the unconstrained broadening processes.

   \subsection{Non-thermal Processes: Global approach}

Here we consider two distinct sources of energetic atoms, depending on the region where the particles are present. For instance, we assume that all energetic atoms produced inside the Roche lobe limit form the so-called internal source, while those produced outside represent the external source. In the frame of our description of the exoplanet's atmosphere (e.g., Figure 4), the internal source could be present in regions I and II, while the external source is confined to region III. 

Generally, the existence of energetic particles results from the departure of their velocity distribution function from equilibrium due to the imbalance between perturbing and restoring processes (\cite{mey01}). For neutrals, the sources of energetic atoms are chemical- and radiation-induced reactions and/or energization by atmospheric meso-turbulence. For partially ionized plasma, the mechanisms of energization could originate from wave-particle interactions or mesoscale turbulence. The resulting velocity distribution varies from one constituent to another, greatly enhancing the broadening effect for particles of larger mass (\cite{shi04, shi07}). For reference, this mass dependence corresponds to the so-called preferential ion heating and acceleration with respect to hydrogen in the extended solar corona (\cite{cra08}). The problem is complex and requires solving Fokker-Planck equations for the particles' velocity distribution function that is sensitive to the neutral or charged state of the background gas (\cite{shi04, shi07}. For example, \cite{shi07} evaluated the mass dependence of the velocity distribution function to wave-particles' interactions that may be quite different from a Maxwellian or even a kappa distribution. Meso-turbulence may also broaden absorption lines in a very complex way, particularly when the gas departs from a steady state (\cite{lou02}). 

Our intent is not to conduct such heavy modeling, but rather to follow the rich heritage of the many studies of the atmosphere of solar system's planets and the heliosphere (\cite{shi96, she04, fah07, nag08, zan09}, and references therein). Energetic atoms are commonly detected in these objects at all distances from the Sun. When we recall the huge stellar energy available at the orbit of giant explanets, we are not persuaded that a sound analysis can neglect or ignore such energetic atoms in their extended atmospheres. Similar to prior studies conducted on the solar corona (\cite{cra98,cra08}) or the Jovian corona (\cite{ben93, eme96, gla04}), the preferential broadening for the OI and CII lines is accounted for by assuming synthetic velocity distributions with one parameter corresponding to the effective temperature $T_i$ of the species ${\it i}$ that should be derived from the comparison to the transit absorption observed for the selected constituent.

 \subsubsection{Impact of internal Source of Energetic Atoms}

In our global approach, we start from the DIV1 standard model of \cite{mun07}, but adopt values of $f_{sca}$ (1, 2/3, 1/2, 1/3, 1/10) for scaling OI and CII abundances in the atmosphere. We do not consider larger values of $f_{sca}$  as models with $f_{sca}\sim 2/3-1.0$ already correspond to an enriched atmosphere and fit rather well with the transit absorption at HI \lya\ (see Figure 7A). Independent scaling of each species may help to locate any departure from solar mixing ratios that are assumed deep in the atmosphere.

The region where collisional and wave-particles' broadening is acting is placed on top of the atmosphere and defined by the bottom position, the first free parameter. In such an active region, the velocity distribution of an individual constituent is allowed to depart from the background. The departure from equilibrium is defined by the ratio $T_i/T_B$ of the specie's effective temperature to the background temperature. The results of the comparison of our model to the observed transit absorption are provided as a function of the atmospheric scaling factor adopted and the position of the bottom of the active layer. Our results are summarized in Table 4 and are discussed below to derive key properties of hot populations produced in regions I and II. 

\begin{table*}
\tablenum{4}
\caption{Summary of sensitivity study versus atmospheric scaling factor and position of hot atoms sources  required to fit HST/STIS archival FUV observations (see section 4). Cases for which thermal and non-thermal line broadening could not explain assumed observations are indicated by "no" for the corresponding specie. Hot atoms sources are confined above the indicated R$_s$ position, where $R_p$ is the planetary radius. The first five positions are for internal sources (region I and II), while the last is for an external source (region III).  } 
\begin{center}
\begin{tabular}{{r}lccccc}
\hline
\hline
 R$_s$  & $\sim 2.26R_p $ & Internal Source  & & & & \\
\hline
   & ($ f_{sca} $)              &      1/10    &     1/3            &     1/2     &      2/3   &   1    \\
    & HI ($\chi^2$; T$_{HI}$/T$_B$) &  ($>3.5$; 9.)  &  (2.07; 2.75) &    (1.93; 2.5)       &  (1.72; 1) &   (1.97; 1) \\        
  & OI (T$_{OI}$/T$_B$)           &  32          &    13.7           &     11.75   &   10.75       &   9.5        \\
  & CII (T$_{CII}$/T$_B$)         &    12.25     &    7.0            &     6.0     &    5.5        &   5.0         \\
\hline
\hline
 R$_s$  & $\sim 2.57R_p $ &  & & & & \\
\hline
 &       $f_{sca}$                   &      1/10    &     1/3       &     1/2     &      2/3   &   1    \\
 &	HI ($\chi^2$; T$_{HI}$/T$_B$) &  ($>3.5$; 10.)  &  (2.18; 2.75) &    (1.93; 2.5)      &  (1.72; 1) &   (1.97; 1) \\        
 &	OI (T$_{OI}$/T$_B$)           &  284          &    20.75     &    16.25   &   14.0    &   12.0        \\
 &	CII (T$_{CII}$/T$_B$)         &    21.5     &    9.25        &     7.75     &    7.0     &  6.0  \\
\hline
\hline
 R$_s$  & $\sim 2.89R_p $ &  & & & & \\
\hline
 & $f_{sca}$                   &      1/10    &     1/3       &     1/2     &      2/3   &   1    \\  
 &	HI ($\chi^2$; T$_{HI}$/T$_B$) &  ($>3.5$; 12.5)&  (2.33; 3.5)  &    (1.97; 3.25)       &  (1.72; 1) &   (1. 97; 1) \\        
 &	OI (T$_{OI}$/T$_B$)           &  no          &    no         &     no      &   94.0     &   38.25         \\
 &	CII (T$_{CII}$/T$_B$)         &    no        &    33.5       &     20.75      &    16.0    &   12.0        \\
\hline
\hline
 R$_s$  & $\sim 3.32R_p $ &  & & & & \\
\hline
 & $f_{sca}$                   &      1/10    &     1/3       &     1/2      &      2/3   &   1    \\
 &	HI ($\chi^2$; T$_{HI}$/T$_B$) &  ($>3.5$; 13.5)  &  (2.35; 4.0) &    (2.0; 3.5)       &  (1.72; 1) &   (1.97; 1) \\        
 &	OI (T$_{OI}$/T$_B$)           &   no           &    no       &     no       &   no       &   369.5        \\
 &	CII (T$_{CII}$/T$_B$)         &    no        &    107.0       &    42.0       &   28.0     &   18.25         \\
\hline
\hline
 R$_s$  & $\sim 3.63R_p $ &  & & & & \\
\hline
 & $f_{sca}$                   &      1/10    &     1/3       &     1/2     &      2/3   &   1    \\
 &	HI ($\chi^2$; T$_{HI}$/T$_B$) &  ($>3.5$; 15.)  &  (2.38; 4.25) &    (2.0; 4.0)       &  (1.72; 1) &   (1.97; 1) \\        
 &	OI (T$_{OI}$/T$_B$)           &   no          &    no       &     no     &   no       &   no         \\
 &	CII (T$_{CII}$/T$_B$)         &   no         &    no        &     no     &   93.0     &   39.0         \\
\hline
\hline
 R$_s$  & $> R_{L1}$ & External Source & & & & \\
\hline
 & $f_{sca}$                  &      1/10    &     1/3       &     1/2     &      2/3   &   1    \\
 & HI ($\chi^2$; T$_{HI}$/T$_B$) &  (1.52; 102.)  &  (1.62; 93.5) &  (1.67; 91.5)       &  (1.72; 1) &   (1.97; 1) \\ 
  &		$[HI]_{hot}$ ($10^{13} {\rm cm^{-2}}$) &   $4.0$ & $2.42$   & $1.3$   & $0.0$     & $0.0$  \\  
\hline
\hline
\end{tabular}   
\end{center}
\end{table*}

First, a quick survey of Table 4 shows that a strong differential heating of OI and CII is required with respect to the background atmosphere in order to fit the assumed OI and CII transit absorptions. This result is certain because preferential atomic heating is required for OI and CII in all possible atmospheric cases in order to recover the transit absorptions observed. In addition, the derived differential heating is here found to be stronger for OI than for CII, which nicely follows the mass dependence as derived by theoretical studies \cite{shi07} and observed in the solar corona (\cite{cra08}). However, the corresponding absorption line broadening depends on the assumed thickness of the non-thermal region and on the scaling factor of the abundance of the species considered over the whole atmosphere. Obviously, our set of observations cannot unambiguously discriminate between the different solutions, yet all of them show a need for a preferential heating of OI and CII compared to the background atmosphere (e. g., Table 4). 

To illustrate the degeneracy of solutions, we consider the HI ($1215.67$\AA), OI ($1304.86$\AA), and CII ($1334.53$\AA) absorption profiles used to fit the transit absorption with our standard atmosphere scaled by a factor $f_{sca}\sim2/3$. For HI, no internal hot hydrogen is required because thermal broadening is enough to bring the model to the desired transit absorption at \lya\ (e.g., Figure 7A). Assuming a hot layer confined to a region above $r\sim 213\times 10^3$ km ($\sim 2.25\, R_p$), a differential heating $T_{OI}/T_{B}\sim 10.75$ is required for OI in order to fit the transit absorption observed around $\sim 1304$\AA, while for CII, $T_{CII}/T_{B}\sim 5.5$ is needed to recover the transit absorption assumed around $\sim 1334$\AA. Now, for depleted atmospheres corresponding to smaller scaling factors and for the same position of the active layer, more efficient preferential broadening is required for OI and CII lines to balance the loss in absorption due to smaller atomic opacity. As shown in Figure 8, for an atmospheric scaling factor ranging from $2/3$ down to $1/10$, $T_{OI}/T_{B}$ varies from $\sim 10.75$ to $\sim 32$ for oxygen, while for CII, $T_{CII}/T_{B}$ increases smoothly from $\sim 5.5$ up to $\sim 12.25$. In other words, the smaller the atmospheric scaling, the stronger the non-thermal broadening needed to fit the observed transit absorptions. Intuitively, this result simply reflects the need for an extra line opacity to compensate for reduced atomic opacity, an extra opacity that is produced by a larger effective temperature. For HI, the fit obtained with an internal source of hot HI is rather poor compared to the fit obtained for thermal broadening (see Figure 7B). Moreover, the corresponding \lya\ spectral shape has a pronounced U-shape that seems different from observations with $\chi^2 > 2$ in all cases, making this process an improbable explanation for the observed transit absorption at \lya\ . Note, however, that if the internal hot atoms are placed high in the atmosphere, close to the Roche lobe edge, then it would be difficult to distinguish them from the external source, a scenario discussed below. In summary, while a standard model with $\sim 2/3$ abundances gives the best $\chi^2$ fit for HI, along with a reasonable picture of the differential line broadening between HI, OI, and CII (e.g., Figure 8), other models with depleted OI and CII abundances (compared to the reference DIV1 model) provide satisfactory fits to observations but with rather stronger broadening needed for OI and CII absorption lines (see Figs. 8 and 9). Later, we show how an external source of hot atoms may modify this picture for HI but not for OI and CII.

\begin{figure}
\epsscale{1.1}
\plotone{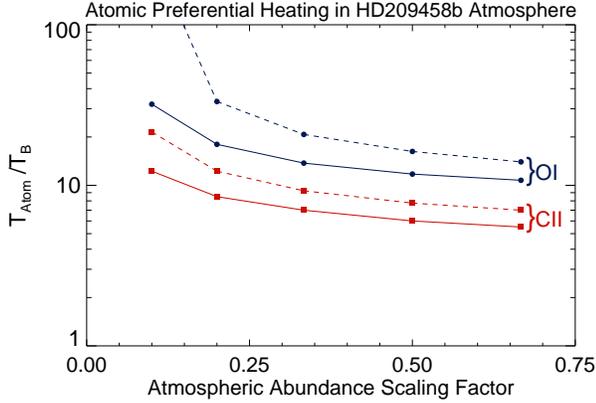}
\caption{Ratio of effective temperature to background gas temperature ($T_B$) for the two species considered, OI and CII, as a function of the atmospheric abundance scaling factor. The scaling factor is applied to the whole atmosphere taken as the DIV1 model from \cite{mun07}. (Solid line) Energetic atoms confined to region above $213\times10^3$km ($\sim 2.25\, R_p$), a position almost $6\times10^4$km below $R_{L''1}$. (Dashed line) same but for a region confined to a position $3\times10^4$km above $R_{L''1}$. Changing the bottom of the hot atoms region not only shifts the whole effective temperature distribution but also may make the region so optically thin for particular values of $f_{sca}$ that no effective temperature could produce the line broadening required to fit the transit absorption. In the example shown, OI has no hot source solution for $f_{sca}=1/10$ and a source position above $\sim 2.57 R_p$.  \label{fig8}}
\end{figure}

After the atmospheric scaling factor, the second important parameter is the size of the atmospheric active layer, where non-thermal line broadening is operating preferentially on OI and CII lines. To explore the sensitivity of the effective temperature to the thickness of the active layer (or, equivalently, to the position $R_s$ of its bottom), we vary the latter position above and below $R_{L''1}$ of the bottom boundary of the bulge region (e.g., Figure 4 for the atmospheric structure). Assuming our standard atmospheric model with $f_{sca}\sim2/3$, the transit absorption profiles obtained for an active layer positioned above $R_s=213\times 10^3$km ($\sim 2.25R_p$), $243\times 10^3$km ($\sim 2.57R_p$) , and $273\times 10^3$km ($\sim 2.89R_p$) are shown in Figures 9A and 9B, respectively, for OI and CII lines. For reference, a line profile with the only case of thermal broadening is also shown. It is interesting to note distinct features in the wings of the line profiles due to the change of the position $R_s$. Indeed, when hot atoms are confined high in the atmosphere, the absorption profile bears two distinct signatures, one narrow distribution from the thermal component around the line core spectral range and a second, more extended distribution in the far wings that originates from hot atoms. It is interesting to note that the jump from one distribution to another appears as a inflection in the absorption line profile that may be confused with a Doppler shifted spectral signature from a putative fast moving population around $\sim \pm 100$km/s for HI and $\sim \pm 20$km/s for OI and CII. The  extra-spectral broadening induced by the change in the hot source position also translates into a shift of the effective temperature as shown in Figure 8 (dashed lines). In any case, the position of the active layer for either OI or CII could not be very high; otherwise, the required heating would be excessive (e. g., Table 4). From another side, it is difficult to maintain a hot layer very deep in the atmosphere because restoring processes become important {with increasing pressures}, making it unlikely that any forcing process can keep a large differential heating between constituents {in a stratified atmosphere}. 

\begin{figure}
\epsscale{1.1}
\plotone{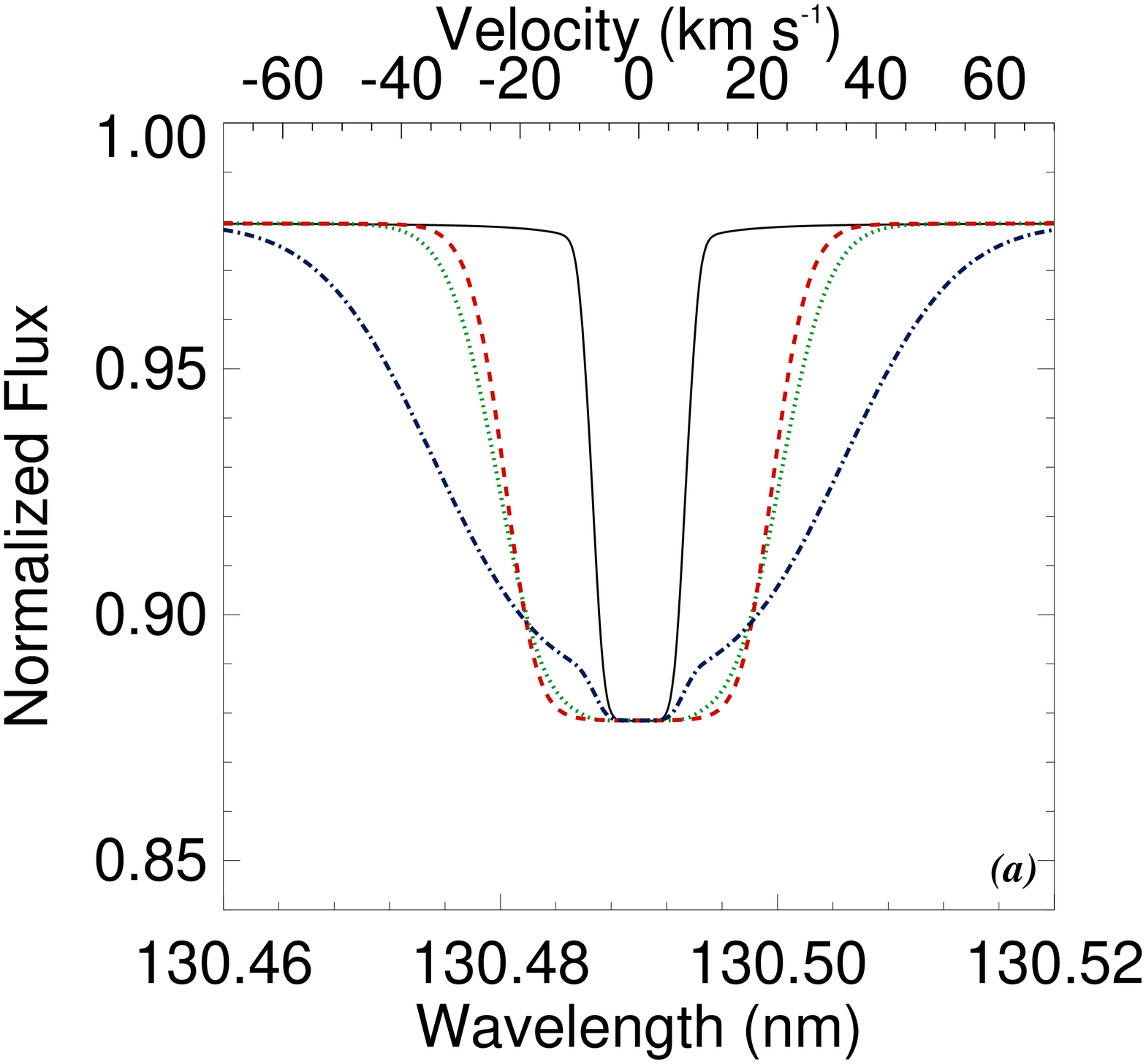}
\plotone{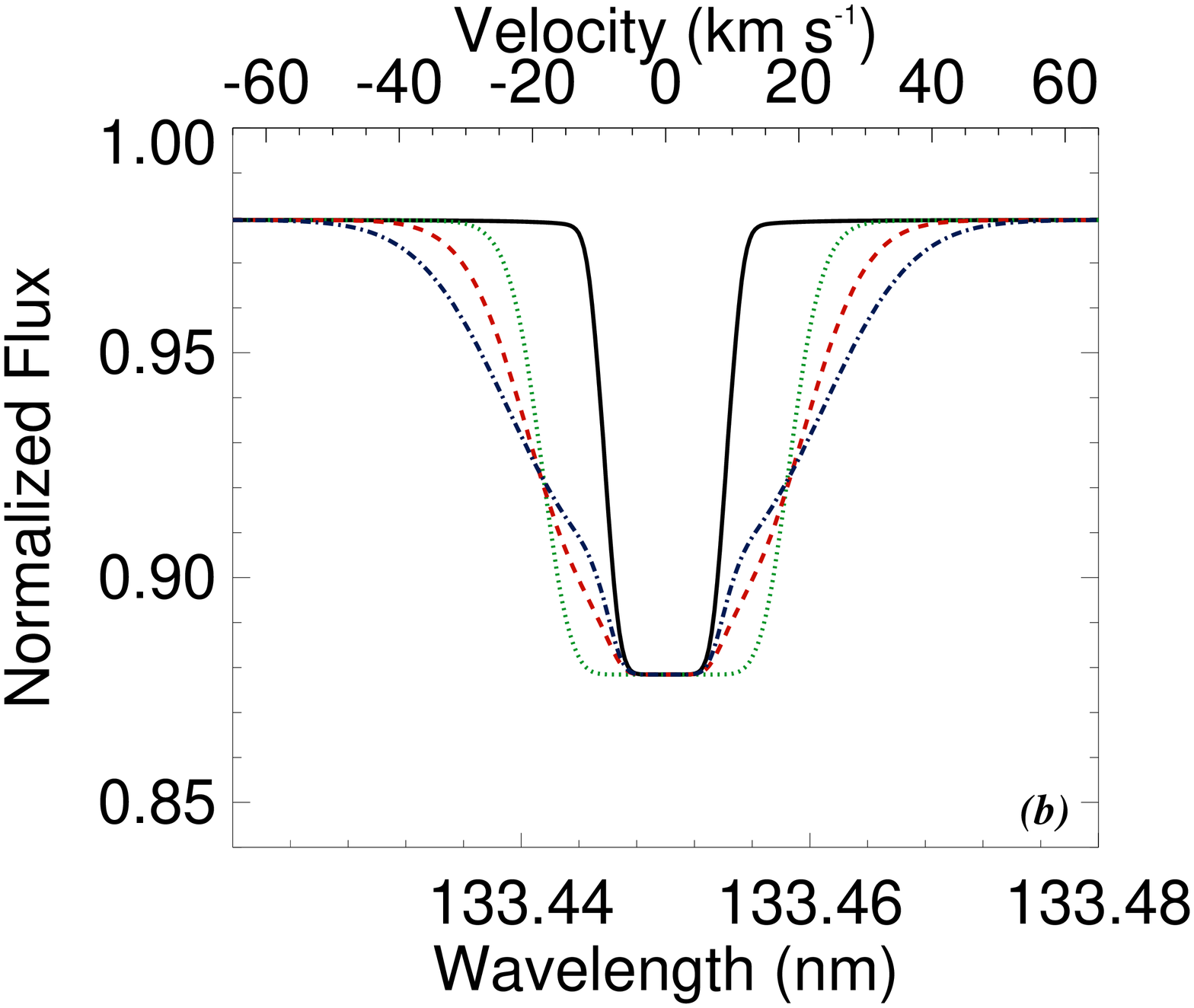}
\caption{(a) Model of OI ($130.486$ nm) absorption profile during full transit estimated for different positions of the hot OI atomic source for an atmosphere scaled from the DIV1 model by a factor of $1/3$. For reference, the solid line shows the transit absorption profile obtained with the same atmospheric model but without the inclusion of hot atomic OI. The dashed line shows a hot OI source confined above $\sim 2.25\, R_p$ and an effective temperature of $T_{OI}/T_B\sim 13.75$; the dotted line shows a hot OI source confined above $\sim 2.57 R_p $ and an effective temperature of $T_{OI}/T_B\sim 20.75$; and the dotted-dashed line shows a hot OI source confined above $ \sim 2.83 R_p$ and an effective temperature of $T_{OI}/T_B\sim 76$. (b) Model of CII ($133.453$ nm) absorption profile during full transit estimated for different positions of the hot CII atomic source for an atmosphere scaled from the DIV1 model by a factor of $2/3$. For reference, the solid line shows the transit absorption profile obtained with the same atmospheric model but with or without the inclusion of hot atomic CII. The dotted line shows a hot CII source confined above $\sim 2.57 R_p$ and an effective temperature of $T_{CII}/T_B\sim 7$, (dashed line) Hot CII source confined above $\sim 2.89 R_p$ and an effective temperature of $T_{CII}/T_B\sim 16$, and (dotted-dashed line) Hot CII source confined above $\sim 3.21 R_p$ and an effective temperature of $T_{CII}/T_B\sim 28$. { Note that for both OI and CII absorption profiles, a spectral inflection appears at $\sim \pm 20$km/s in the wings when hot atoms are included}.  \label{fig9}}
\end{figure}

\subsubsection{Impact of External Source of Energetic atoms}

The last parameter of importance in our study is the impact of an external hot atomic source. First, we recall that the spectral signature of such a population should show little to no Doppler shift to be consistent with the medium-resolution \lya\ observation (\bjlb). Second, we recall our initial conclusion that the atmospheric scaling factor should not exceed $2/3$, a value for which no hot population is required to recover the observed transit absorptions at \lya\ (e. g., Section 4.1). Third, the origin of the external source could be stellar or populations lost from the planet over time. With that in mind, five values of the atmospheric scaling factor $f_{sca}$ are considered in the range $1/10$-$2/3$. Then, for each value of $f_{sca}$, we vary the thickness and temperature of the external hot layer as free parameters in order to minimize the $\chi^2$ corresponding to the \lya\ observations. For all values of $f_{sca}$, our study shows that having an additional, external layer of hot HI atoms yields transit absorptions that compare well with flux-integrated and line profile \lya\ observations but with rather different spectral signatures that could not be constrained from the only data available. As shown in Figure 7B, the external source, the internal source, and the source-free cases show distinct spectral signatures in the absorption profile during transit that should help to discriminate between the different solutions when high-resolution and better \sn\ $\,$ observations become available. Despite this degeneracy, our study shows that if the atmospheric scaling factor $f_{sca}$ is strictly smaller than $2/3$, a population of hot HI atoms is then required. If this population is confined on top of the Roche lobe limit, its parameters (hot HI column and effective temperature) should follow the distribution shown in Figure 10. Interestingly, for most solutions, the effective temperature seems to fall around $T_{HI}/T_B \sim 90-110$ (or $T_{HI}\sim 1.2\times 10^6$K for $ T_B=1.2\times 10^4$K) while the hot HI column is in the range $[HI]_{hot}\sim \left(2-4\right)\times 10^{13} {\rm cm^{-2}}$. It is remarkable to see that, if assumed of stellar origin, the external HI source derived here compares nicely with the ENA  solution proposed by \cite{hol08}, the only difference being that their bulk speed should be weak. 

\begin{figure}
\epsscale{1.1}
\plotone{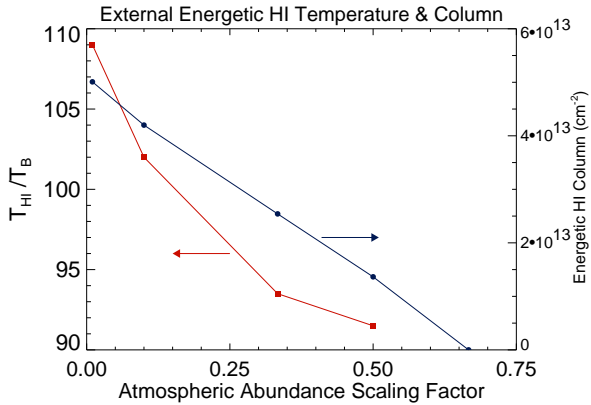}
\caption{External hot HI source (region III) parameters' ($[HI]_{hot}$, $T_{HI}/T_B$) sensitivity to atmospheric scaling factor of ($f_{sca}$). For each value of $f_{sca}$, the hot HI layer parameters are derived from the best fit to \lya\ observations in terms of $\chi^2$. For $f_{sca}=2/3$, no hot HI atoms are required.   \label{fig10}}
\end{figure}

By contrast, for OI and CII lines, our results show that the transit absorption is insensitive to the presence of hot atoms in region III. This non-sensitivity could be explained by the fact that region III is optically transparent for all UV lines of heavy elements considered here because the corresponding abundances are greatly depleted near the Roche lobe limit ($\sim10^3-10^4 {\rm cm^{-3}}$). For these reasons, such an external population, if it exists, could not on its own explain the strong absorption observed for OI and CII unless an additional non-thermal line broadening is included yet is deeper inside regions I and II for the two constituents (see the previous section).

\section{Exoplanet-Star Environment: Role of Energetic atoms}

At this level, it may be worth casting our result in the general context of a giant exoplanet's interaction with the harsh environment caused by its nearby parent star. In that context, the roles of the stellar wind and its related magnetic field are not clear, particularly in regard to their merging with the planetary flow and magnetic field. Assuming that hot planets are locked with the same side facing the star, some studies derived that, contrary to our quickly-rotating Jupiter, the strength of their magnetic field should be very weak, or it may not exist at all (\cite{ip04, san04, zar07, gri07}). In that context, hot atomic populations should be produced in the system with an energy spectrum and a spatial distribution that should both depend critically on the assumed magnetic configuration in the plasma. As illustrated by comparable situations in our solar system such as the interaction between the solar wind and the LISM, or the interaction between the Jovian magnetospheric plasma and the outflow from the satellite Io, one may legitimately postulate that detection of hot atoms from an exoplanetary system should reveal new pieces of information regarding the spatial distribution of the plasmas and magnetic fields present and the strength of the interaction processes in play (\cite{ben00, rat02, zan09, com98}). While the diagnostic will not be easy, we believe it is fundamental for deriving key parameters of the star-planet magnetic environment. It follows, then, that a self-consistent modeling of the interacting planet-star system, though difficult, will be much-needed in the near future. The final configuration will be complex and will surely require three-dimensional MHD modeling of the interacting plasmas.

In our most recent study, we discussed most of the existing models proposed to describe the gas distribution in the HD209458b system based on HST \lya\ observations (\bjlb). More recently, one-dimensional and two-dimensional hydrodynamic simulations have been proposed in the literature and draw the interesting conclusion that the stellar wind may, under specific conditions, confine the planetary flow and thereby possibly enhance the total opacity during transit (\cite{mur09, sto09}). This possibility offers an additional, intriguing process for enhancing the system opacity from outside compared to the internal enhancement induced by heavy elements chemistry as described in \cite{mun07}. Such a possibility could prove to be of no small importance in uncovering the balance between thermal and non-thermal processes in shaping the transit absorption for all species, particularly for OI and CII. 

The problem with most models proposed in the literature is that they fail to recover the observed \lya\ transit absorption. For instance, starting from a one-dimensional hydrodynamic description of the interaction of a planetary wind with a stellar wind but missing the atmospheric chemistry, \cite{mur09} derived a transit absorption, corresponding to the wavelength range defined by \vma\ and \bjlb, of $\sim 2.9\%$ that falls short of the $8.4\%-8.9\%$ absorption reported in \bjlb. These authors thus questioned the accuracy of the transit absorption derived at this point from the HST observations, arguing that the transit \lya\ absorption reported from other low-resolution observations was much weaker (\vmb). Modeling the fully-integrated \lya\ line absorption during transit, they thus predicted $\sim 2.4\%$ absorption that still falls short of the absorption level reported thus far. Within that frame of reasoning, \cite{mur09} thus concluded that the origin of the missing absorption in their wind model at $\pm100{\rm km s^{-1}}$ from line center was uncertain and that a good candidate could be the charge exchange between the planetary and stellar winds. However, we recall that an absorption at the "magic number" of $\pm100{\rm km s^{-1}}$, repeatedly discussed in several studies since the first report of the HST observations in 2003, could be the signature of a population flowing at that velocity range, yet it could also be an opacity effect purely related to a simple curve of growth broadening of an absorption line by the gas opacity (\bjlb). For reference, the wavelength range reported in \vma\ and \bjlb\ was virtually dictated by the wavelength range where the HST spectrum has a good \sn\ $\, $ and is not truly directly related to the observed object. Secondly, as we show here (see Section $3$), the low- and medium-resolution observations are now compatible and there is no opposition between the transit absorption line profile shown in Figure 1 and the transit absorption reported for the integrated line (e. g., Table 2). It follows that if the \cite{mur09} predictions for the transit \lya\ absorption fall short of the observations, respectively, from medium- and low-resolutions, this only means that their model does not render an accurate picture of the gas distribution in the HD209458 system. As a matter of fact, when scaling the hydrogen content in their model by a factor of $3-5$ as proposed by \bjlb, these authors could obtain a transit absorption comparable to observations. However, invoking charge exchange without any restriction is probably not a good idea if the stellar wind speed is strong enough to leave a spectral signature that is not observed (\bjlb). 

In contrast to models previously described, here we take a forward analysis approach that proves very useful, at least for translating the transit absorptions observed for different species in the HD209458 system into atmospheric parameters and processes. First, our analysis of the FUV HST/STIS archives confirms relatively strong absorptions previously reported for OI and CII but with larger statistical errors.  For the heavy elements, our analysis shows that a preferential heating is required, which is consistent with the energy balance in play due to the proximity of the parent star, either through tidal disturbances or the impact of the stellar radiation field and the stellar wind particles and magnetic field. {Indeed, hot atoms are minor populations that receive energy from both the atmospheric pool of the HI dominant species and the stellar impinging particles and photons. The two sources are connected but have enough energy to over-heat minor components  thanks to the imbalance between perturbing and restoring processes (\cite{mey01}). For example, assuming only 1\% of the thermal energy stored in the background gas (at the background temperature $T_B\sim 10^4$ K) is transmitted via non-equilibrium processes to minor constituents (mixing ratio $\sim 10^{-4}$), it is not difficult to find that minor elements could be over-heated by large factors that may explain the temperatures reported here}. In addition, the large line broadening required to recover the observed transit absorptions in all lines of heavy species is consistent with the theoretical prediction of preferential heating of heavy species in a non-equilibrium gas. For OI, the $\sim 2.3\sigma$ transit absorption detection requires hot OI with an effective temperature ratio of $T_{OI}/T_B\sim 10$ or more, depending on the thickness of the active region. For CII, the transit absorption comes with larger statistical errors, yet it also requires hot CII ions with an effective temperature ratio of $T_{CII}/T_B \sim 5$ or more, but definitely smaller than the temperature ratio obtained for OI. As shown in Figure 8, the ratio between $T_{OI}$ and $T_{CII}$ seems consistent with theoretical predictions about the mass dependence of the broadening velocity distribution of heavy elements in disturbed plasmas (\cite{shi07}). Conversely, starting from such theoretical models, if non-thermal heating is detected for the OI atoms (with a mass $M=16$), as here, then we must expect similar heating of the CII ions (with a mass $M=12$), but not as strong as that obtained for OI. With that in mind, we may confidently state that both observational and theoretical results tend to support a scenario of preferential heating of OI and CII with respect to the background atmosphere of HD209458b. The answer to exactly how much and where atomic energization (or non-thermal heating) is operating will require future observations with better \sn\ $\,$ and spectral resolution. 

For the dominant component HI, the observations are better quality, yet the conclusions are still uncertain. Our sensitivity study favors two families of solutions for HI (see section 4). The first family requires only thermal HI in a thick atmosphere with a large dayside vertical column, while the second family is based on HI-depleted atmospheres with respect to the DIV1 model (depletion larger than $3/2$), but includes a thin source of hot HI confined to the outer layers of the Roche lobe limit of the nebula. If the second solution is confirmed (see Figure 7), high-resolution and \sn\ $\,$ \lya\ absorption profiles observed from ingress to egress orbital positions should provide the related orbital distribution of energetic neutral HI (effective temperature and HI column), which in turn should help map the spatial distribution of the energization processes in play in the system. In addition, if the hot HI source is of stellar origin, our results place strong constraints on the stellar wind and should require it to be rather slow, consistent with the picture of the exoplanet evolving in the stellar extended corona where the flow still may be slow compared to farther distances (\cite{pre05, gri07}). Finally, we note that even if the external source appears to be a potential option for HI, it is not a viable solution for the less-abundant heavy species OI and CII, for which an energization process is required deeper in the atmosphere.

At this stage, one may wonder what diagnostic can assist in obtaining the thickness and effective temperature of the regions where the different energetic atoms are confined. First, we note that the preferential heating described here should not be uniformly distributed in a confined region, but rather should be distributed with altitude. Now, depending on the depth at which hot atoms are present, our model calculations seem to indicate that all HI, OI, and CII absorption line profiles predicted during full transit should show distinct shapes, particularly in the line wings (see Figures 7 and 9). This means that high-resolution and \sn\ $\,$ observations obtained for distinct species (the mass effect) may help discriminate between the different solutions. Of particular interest, the OI and CII transit absorption profiles should not only constrain the balance between thermal and non-thermal atomic populations for these species, but also help to answer the question of whether heavy elements in hot Jupiters are depleted with respect to solar abundance values. In addition, ingress and egress observations with high \sn\ $\,$ should provide more details about the spatial distribution of the occulting plasma, which in turn should bring new insight as to the magnetic configuration and energization processes via the inferred abundance and distribution of the energetic populations. 

\section{ Summary and Conclusions} 

HST archive low- and medium-resolution observations of the HD209458 system in the FUV spectral domain are revisited to uncover the gas distribution and kinetic processes in the nebula that envelops the exoplanet. Within that framework, three important constituents-HI,  OI, and CII-are considered. Our analysis of the low-resolution observations obtained in the wavelength range of $1180-1710$ \AA\ shows a transit absorption of $\sim 6.6\%\pm 2.3\%$ for HI, $\sim 10.5\%\pm 4.4\%$ for OI, and $\sim 7.4\%\pm 4.7\%$ for CII. Past results from the analysis of medium-resolution observations for the HI \lya\ component are utilized here (\bjlb). Starting from a standard model that includes most of the important chemistry in play in an atmosphere forced by radiation, particles, and gravity from the star, we built a more sophisticated model to account for broadening processes that may originate from superthermal atoms in the upper layers of the nebula or that may be of stellar origin. To estimate the transit effect, the stellar line profiles must be known. While the stellar \lya\ line profile is known to a high degree of accuracy from HST medium- and high- spectral resolution observations, such is not the case for the OI and CII lines for which existing solar line profiles are assumed. The major difference between this study and previous ones is consideration of OI and CII lines simultaneously with the HI \lya\ line. The challenge was to obtain for OI and CII almost the same level of transit absorption as for HI when the abundance is at least 3 or more orders of magnitude smaller for the heavy elements, particularly when the atomic absorption cross-sections for the three sets of lines have the same order of magnitude. A differential process that must compensate for the OI/HI and CII/HI lower abundances is then required in order to determine the comparable magnitude of transit absorption by the three species. Here, we propose particles-particles collisions, wave-particles interaction, and mesoturbulence effects that may enhance line broadening during the stellar radiation transmission. Such processes are driven by huge forces caused by the impact of stellar radiation, impinging particles, and tidal and magnetic distortions, all of which act cumulatively on the exoplanetary atmosphere.

Three families of models are thus considered: first, a case free of energetic atoms; second, a case with hot atoms included internally (inside regions 1 and 2); and third, a case with external hot atoms (within region III). With an independent scaling factor for each specie, this description covers most models discussed or proposed to date in the literature. First, the comparison of our theoretical model to HST observations of HD209458 shows that for all models considered, a population of hot OI and CII atoms, confined inside the Roche lobe, is required in order to fit with the transit absorption observed for the two species' lines. Our parametric analysis clearly shows that for all considered cases, a preferential heating of OI and CII compared to HI is required, with a magnitude that depends on the assumed atmospheric model (e.g., Figures 8 and 9) but with an effective temperature that is definitely higher than the background temperature (e.g., Figure 8). In addition, all effective temperatures derived here are higher for OI than for CII, consistent with the theoretical prediction of the mass dependence of velocity distribution of heavy elements in a non-equilibrium plasma (\cite{shi07}). Interestingly, the presence of energetic atoms may translate into a spectral inflection that appears $\sim \pm 20$km/s from the line center of the OI and CII lines (Figure 9). Future detection of this spectral feature would confirm the direct signature of superthermal atoms populating the upper atmosphere of HD209458b. 

For HI, by contrast, two families of models provide a satisfactory fit to available \lya\ observations (see Figure 7). For instance, when considering a model free of energetic HI atoms, it is possible to fit both the transit absorption profile and the integrated flux drop-off at \lya\ simultaneously with an atmosphere that has a dayside column $[HI]\sim 1.05\times 10^{21}{\rm cm^{-2}}$, corresponding to a depletion in HI by a factor $3/2$ with respect to the DIV1 model of \cite{mun07}. However, if we assume a more depleted HI atmosphere with respect to DIV1 model, then a population of energetic HI atoms is required in order to fit the observations (e.g., Figures 7 and 10). Our sensitivity study shows that depending on the hot source position, which may be confined inside or outside the Roche limit, distinct spectral features appear at $\sim \pm 100$km/s in the wings of the transit absorption profile that only future high-resolution observations with a decent \sn\ $\,$ should be able to detect (e.g., Figure 7). Among the two potential sources of hot HI atoms that may explain the \lya\ observations during transit, the one confined near or outside the planetary Roche lobe is of particular interest because it may be directly related to the stellar wind. Indeed, if this hot HI population (region III) is of stellar origin, its parameters may help constrain the stellar wind speed and flux. In most HI-depleted models considered here, the required external source has an HI column in the range of $[HI]_{hot}\sim \left(2-4\right)\times 10^{13} {\rm cm^{-2}}$ and an effective temperature in the range of $T_{HI}/T_B \sim 90-110$ (see Figure 10). 

In summary, our sensitivity study shows that multi-species observations including HI and heavy elements are the key tools essential for understanding the structure and composition of transiting exoplanets and their interaction with the nearby parent star. With all predictions expounded here, we look forward to the upcoming HST/Cosmic Origins Spectrometer (HST/COS) and STIS observations that should finally reveal the true balance between thermal and non-thermal populations and, we hope, their spatial distribution.

\acknowledgments
L. B. J. acknowledges support from Universit\'e Pierre et Marie Curie (UPMC) and the Centre National de la Recherche Scientifique (CNRS) in France. The authors warmly thank Dr. I. Hubeny and Dr. R. Gray for helpful discussions on HD209458 stellar spectra. Part of this work was done during the first author's visit to U. C. Davis in 2008. L. B. J. warmly thanks the faculty and technical staff at the Applied Science Department at U. C. Davis for their hospitality, particularly Professor W. Harris and Mrs. T. Macias. S. H. contributed to this work during her graduate training program in the Department of Applied Science. This work is based on observations with the NASA/ESA Hubble Space Telescope and obtained at the Space Telescope Science Institute operated by AURA, Inc. Authors acknowledge support from the University of California, Davis.

\end{document}